\documentclass[12pt,a4paper]{article}
\pdfoutput=1

\usepackage[left=2.5cm,top=2.5cm,right=2cm,nohead]{geometry}
\usepackage[colorlinks=false,
   linkcolor=red, 
   citecolor=blue,
    filecolor=red,
    urlcolor=red,
    linktoc=all, 
    pdfstartview=FitV,
    bookmarksopen=true]{hyperref}
\usepackage[utf8]{inputenc}

\usepackage{amssymb}
\usepackage{tikz}
\usetikzlibrary{cd}
\usetikzlibrary{babel}
\usepackage{verbatim}
\usetikzlibrary{matrix}
\usepackage{adjustbox}
\usepackage[titletoc]{appendix}
\usepackage{graphicx}
\usepackage{float}
\usepackage{url}

\usepackage{subfig}
\usepackage{amsmath}
\usepackage{amsfonts}
\usepackage{mathtools}
\usepackage{amsthm,thmtools}
\usepackage{esint}
\usepackage{wrapfig}
\usepackage{comment}
\usepackage{color}

\usepackage{braket}

\usepackage{ulem}
\usepackage{soul}
\usepackage{centernot}
\usepackage{cancel}
\usepackage{booktabs}

\DeclareMathOperator{\sign}{sign}
\DeclareMathOperator{\tr}{tr}
\DeclareMathOperator{\diag}{diag}


\usepackage{mathtools}

\usepackage{nameref}

\usepackage{cleveref}

\newcommand{\rO}{{\mathrm{O}}}
\newcommand{\be}{\begin{equation}}
\newcommand{\ee}{\end{equation}}

\newcommand{\bea}{\begin{eqnarray}}
\newcommand{\eea}{\end{eqnarray}}
\numberwithin{equation}{section}

\makeatletter

\begin{document}

\begin{center}     
\pagestyle{empty}

  {\bf\LARGE New non-perturbative de Sitter vacua in $\alpha'$-complete cosmology     \\ [5mm]}

\large{ 
Carmen A. N\'u\~nez and Facundo Emanuel Rost
 \\[2mm]}
{\small  Instituto de Astronom\'ia y F\'isica del Espacio (IAFE-CONICET-UBA)\\ [2mm]}
{\small  Departamento de F\'isica, FCEyN, Universidad de Buenos Aires (UBA) \\ [2mm]}
{\small\it Ciudad Universitaria, Pabell\'on 1, 1428 Buenos Aires, Argentina\\ }

{\small \verb"carmen@iafe.uba.ar, facundo.rost@gmail.com"}\\[2cm]

\small{\bf Abstract} \\[7mm]\end{center}
The $\alpha'$-complete  cosmology developed by Hohm and Zwiebach classifies  the $\rO(d,d;{\mathbb R})$ invariant theories involving  metric, $b$-field and dilaton that only depend on time, to all orders  in $\alpha'$. Some of these theories
feature non-perturbative isotropic de Sitter vacua in the string frame,
generated by the infinite number of higher-derivatives of  $\rO(d,d;{\mathbb R})$  multiplets.   Extending the isotropic ansatz, we construct  stable and unstable  non-perturbative de Sitter solutions  in the string and Einstein frames. The generalized equations of motion admit new solutions,  including anisotropic $d+1$-dimensional  metrics and non-vanishing $b$-field. In particular, we find dS$_{n+1}\times T^{d-n}$ geometries with constant dilaton, and also metrics with  bounded scale factors in the  spatial dimensions with non-trivial $b$-field. We discuss the stability and non-perturbative character of the solutions, as well as possible applications.

\newpage 

\tableofcontents

\newpage

\section{Introduction}

The astronomical observations point in the direction that we  live in a flat universe whose expansion is currently undergoing acceleration, in a phase that can be approximated by de Sitter (dS) geometry. In an opposite direction, the construction of a consistent  dS  vacuum in 1+3 non-compact directions in string theory  has proved to be  extremely difficult \cite{vanriet}. Moreover,  a no-go theorem states that there are no (macroscopic) stable or unstable dS$_n$  solutions for $n\ge4$ at tree-level in
 heterotic and type II string theories (in the absence of RR fluxes)
 \cite{kutasov}, and even further, it is conjectured that no stable or meta-stable
dS vacua can
exist in a consistent quantum theory of gravity \cite{swampland}.

Understanding the cosmological consequences of classical string theory  requires the knowledge of 
 the infinitely many higher order  corrections that it  induces on
the Einstein equations. But so far, only the first few lowest orders have been computed explicitly. Since the truncated expansion  may not display the properties of the full string theory,  alternative models for the evolution of the universe
have been constructed from duality invariant    theories with  higher-derivative corrections  to all orders  that are relevant for cosmology \cite{hohm}-\cite{branden2}. 

 The motivation for the so-called $\alpha'$-complete cosmology follows from the observation  by Sen \cite{sen91}  that the string low energy effective field theory
in $d+1$ dimensions displays an $\rO(d,d;\mathbb{R})$  symmetry, also named duality symmetry, to all orders in the inverse string tension $\alpha'$,
when  the fields do not depend on the $d$ spatial coordinates. As discussed in \cite{hsz15}, the dimensionally reduced theory can be obtained from compactification in a $d$-torus $T^d$, ignoring all  Kaluza-Klein excitations that arise from
field configurations in which the fields depend on the compact space.

Using $\rO(d,d;\mathbb R)$   multiplets as field variables,  and assuming that their duality transformations remain unchanged to any order in $\alpha'$, all higher-derivative terms consistent with duality invariance for  purely time-dependent backgrounds were classified in the seminal papers \cite{hohm}   into single- and  multi-trace factors, involving only first derivatives of the fields. This simplification is accomplished by performing duality covariant field redefinitions order by order in $\alpha'$, and assuming that all non-derivative dependence  on the dilaton is contained in the exponential prefactor of the integration measure.
 Implementing an ansatz with a $d+1$ dimensional  isotropic  Friedmann-Lemaitre-Robertson-Walker  metric and   vanishing Kalb-Ramond field, Hohm and Zwiebach showed that  $\alpha'$-complete cosmology  admits non-perturbative dS solutions in the string frame \cite{hohm}.   This type of solutions were also obtained in presence of matter sources in \cite{branden}, while the conditions to get dS vacua in the Einstein frame with non-constant dilaton  were stated
 in \cite{krishnan} in terms of a quite non-trivial second order non-linear differential equation for a function that describes the Lagrangian.

A more realistic cosmological model was constructed in \cite{branden2}, extending
the isotropic ansatz of \cite{hohm} to geometries with two  scale factors: a dynamical one  in $n<d$ spatial dimensions and a constant one in the remaining $d-n$ spatial coordinates.   Inspired by the String Gas Cosmology scenario \cite{branvafa}, the non-perturbative equations of the $\alpha'$-complete cosmology were shown to be (in principle) compatible with a dynamical mechanism  in which  the universe emerges from
a phase with  $\mathbb{R}\times T^9$ 
 spacetime geometry,  with matter made of a gas of strings  and evolves towards  four  large spacetime dimensions  with the
 six internal dimensions stabilized around the string length.

In the first part of this paper, we reconsider the non-isotropic ansatz of \cite{branden2} in the vacuum and rederive the equations of motion. We find the following conditions  to get string frame dS solutions  in $n+1\le d+1$ dimensions when the $\rO(d,d)$ Noether charge vanishes:
\begin{equation}
F_n'(H_0)=0~~,~~F_n(H_0)=c^2\geq 0~~,~~\displaystyle\frac{\partial\Phi}{\partial t}=-c~~,
\end{equation}
where $F_n(H)$  describes the Lagrangian of the theory and is well-defined for non-infinitesimal values of $\sqrt{\alpha'}H$, $H=H_0=$ constant is the Hubble parameter, $\Phi$ is the generalized dilaton, 
and $c$ is a real constant. The configurations described by  $(H_0,c)$ have constant dilaton if
$c=nH_0$
and are dS in both the string and the Einstein frames. Thus the conditions to obtain dS vacua in the Einstein frame are now simple algebraic equations  because they have constant dilaton. These solutions  are stable (unstable) if $c>0$ ($c<0$), and they can be easily classified in expanding or contracting dS geometries.

Both the isotropic and non-isotropic ansatze considered in \cite{hohm}-\cite{branden2} lead to the additional simplification that the multi-trace terms give the same
structural contributions as the single-trace term. However, for more heterogeneous metrics or non-vanishing Kalb-Ramond field, the multi-trace terms cannot be absorbed into single-traces and have to be fully considered. This issue is the subject of the second part of this paper. We include the multi-trace corrections in the equations of motion of the $\rO(d,d)$ invariant cosmology and examine a generalized ansatz in which the metric, its time derivative and the time derivative of the $b$-field are commuting matrices. Although this is a rather general assumption, it turns out that the metric can be made diagonal and both the $b$-field and its derivative can be made block diagonal matrices,  without loss of generality due to the duality symmetry. 

Despite the fact that  the multi-trace corrections cannot be absorbed into single traces, still the field equations can be worked out and allow non-perturbative isotropic and anisotropic dS solutions in $n\leq d$ spatial dimensions. These solutions may have constant dilaton,  thus being stable or unstable dS geometries in both the string and the Einstein frames. The new  vacua are also found in the sector of vanishing $\rO(d,d)$ Noether charge $\mathcal{Q}=\mathbf{0}$, which is then identified as a rich source of non-perturbative dS solutions. 
On the other hand, an interesting effect of the $b$-field  dynamics is that  it must be turned on  in dimensions with non-zero eigenvalues of a Noether charge block (hence in the $\mathcal{Q}\neq\mathbf{0}$ sector), and  the scale factors of such spatial dimensions must be bounded, thus precluding dS solutions.

Besides providing the equivalence between the string and the Einstein frame vacua,  the solutions with constant dilaton $\phi_0$ have the advantage that one can take $g_s=e^{\phi_0}\ll1$. This is consistent with classical string theories, which are some particular points in the space of duality invariant theories. However, these special points may admit non-perturbative dS solutions only if they evade the assumptions of the no-go theorem of \cite{kutasov}. In addition, the string low energy effective  Lagrangian is  an asymptotic expansion in powers of $\alpha'$, and even if all the perturbative contributions were known, we will argue in section \ref{sec:caveats} that non-perturbative information is necessary in order  to be able to assert that the theory admits non-perturbative dS solutions.  In any case, specifying the conditions that  allow dS and other intriguing cosmological solutions in duality invariant
theories  is an interesting result, which may reveal  general features that
apply to string theory.

The paper is organized as follows. In section \ref{sec:alpha}, we present a brief review of $\alpha'$-complete cosmology and  rederive the equations of motion in order to include the multi-trace corrections. Considering an ansatz with one dynamical scale factor  in $n\leq d$ spatial dimensions and a constant one in the remaining $d-n$ dimensions 
 with vanishing $b$-field, in section \ref{sec:feq} we determine the conditions to have dS solutions with constant dilaton. We also analyze the stability of the solutions and classify them. In section \ref{sec:matrixds2} we introduce the generalized ansatz of commuting matrices and work out the corresponding equations of motion, leaving  details of the calculations to  appendix \ref{app:geneom}. We find dS solutions of the new field equations in section \ref{sec:bneq0},  with anisotropic geometries or with non-vanishing $b$-field. These have $\mathcal{Q}=\mathbf{0}$, which is a rich sector for dS solutions, as explained in appendix \ref{app:qne0}.
A summary of the procedure to  follow in order to obtain non-perturbative dS solutions and a discussion of their non-perturbative character is the subject of section \ref{sec:caveats} and appendix \ref{app:caveats}. Finally, an outlook and conclusions are contained in section \ref{sec:conclu}.

\section{\texorpdfstring{$\rO(d,d)$}{1} invariant \texorpdfstring{$\alpha'$}{1}-complete cosmology}\label{sec:alpha}

In this section we briefly review the $\rO(d,d;{\mathbb R})$ invariant  cosmology to all orders in  $\alpha'$  introduced by Hohm and Zwiebach \cite{hohm}, mainly to set the notation. We refer to the original papers for details. In \ref{sec:eom} we generalize  the derivation of the equations of motion in order to include the contributions from the multi-trace corrections, which enable  the construction of non-isotropic solutions and the addition of non-vanishing $b$-field in the forthcoming sections.

\subsection{Field variables  and  action}

 The seminal framework developed in \cite{hohm} is based on Sen's observation  \cite{sen91}  that the low energy effective field theory of the universal gravitational sector of string theory
in $D=d+1$ dimensions displays a global $\rO(d,d;\mathbb{R})$  symmetry to all orders in $\alpha'$, when  the fields do not depend on the $d$ spatial coordinates. This symmetry, also
referred to as ‘duality’, contains the scale-factor duality $a\leftrightarrow a^{-1}$ \cite{vene,vm}. Following \cite{meissner97} and using
 string field theory arguments \cite{kugozw92}, Hohm and Zwiebach 
 assumed that  the $\rO(d,d;\mathbb{R})$  transformations
remain unchanged to all orders in $\alpha'$ if  the theory is expressed in terms of the duality  invariant dilaton $\Phi$,
\be
e^{-\Phi}=\sqrt{\det g_{ij}}~e^{-2\phi}\, ,
\end{equation}
and an $\rO(d,d)$ covariant matrix $\mathcal{S}$, constrained to satisfy $\mathcal{S}^2=1$ and $\mathcal{S}^t=\eta \mathcal{S}\eta$,  where   $\eta=\displaystyle\left(\begin{smallmatrix}0&{\bf 1}\\
{\bf 1}&0
\end{smallmatrix}\right)$ is the $\rO(d,d;\mathbb{R})$ invariant metric.  Every matrix $\mathcal{S}$ that verifies these constraints can be written  in terms of  symmetric and antisymmetric  matrices $g$ and $b$, respectively, as 
\begin{equation}
\mathcal{S}=\begin{pmatrix}bg^{-1}&g-bg^{-1}b\\
g^{-1}&-g^{-1}b
\end{pmatrix}\, ,
\end{equation}
where $g$ and $b$ are the spatial components $g_{ij}$ and $b_{ij}$  of the  space-time metric and Kalb-Ramond fields, that are taken as \footnote{Without loss of generality for fields that only depend on the (non-periodic) time coordinate}: 
\begin{equation}\label{eqgbphi}
g_{\mu\nu}(t)=\begin{pmatrix}-{n}^2(t)&0\\
0&g_{ij}(t)
\end{pmatrix}~\text{with }n(t)>0 ~~~ {\rm and} \quad b_{\mu\nu}(t)=\begin{pmatrix}0&0\\
0&b_{ij}(t)
\end{pmatrix}\, .
\end{equation}

 Further assuming that all non-derivative dependence of the action on the dilaton is contained in the exponential prefactor of the integration measure  and performing duality-covariant field redefinitions,  Hohm and Zwiebach showed that every $\rO(d,d)$  and time reparameterization invariant action describing the dynamics of $\mathcal{S}$ and  $\Phi$  can be brought to the form
\begin{equation}\label{eqactiontotal}
I(\mathcal{S},\Phi,n)=\int dt~n~e^{-\Phi}\left[-(\mathcal{D} \Phi)^2-\mathcal{F(DS)}\right]\, ,
\end{equation}
where $\Phi(t)$ and $\mathcal{S}(t)$ are  scalars under time reparameterization, while $n(t)$ is a density.  Under $h\in \rO(d,d;\mathbb{R})\iff h\eta h^t=\eta$ with $h$ constant, $\mathcal{S}$ transforms  as $\mathcal{S}\rightarrow \mathcal{S}'=h\mathcal{S}h^{-1}$ preserving the constraints, while $\Phi$ and $n$ are  invariant.  The covariant time derivative is
$\mathcal{D}\equiv \displaystyle\frac{1}{n(t)}\frac{\partial}{\partial t}$, and   a time parameterization $t_S$ can always  be chosen such that $n(t_S)=1$.

The function $\mathcal{F(DS)}$ is defined by the following asymptotic expansion
\begin{equation}\label{eqdeffcursiva}
\mathcal{F(DS)}\equiv  -c_1\text{tr}\left[(\mathcal{DS})^2\right]-\displaystyle\sum_{k=2}^\infty~\alpha'\,^{k-1}~\displaystyle\sum_{P\in~\text{Part}(k,2)}c_{k,P}\prod_{m\in P}\text{tr}\left[(\mathcal{DS})^{2m}\right] \, ,
\end{equation}
which contains all the $\alpha'$-corrections. $\text{Part}(k,2)$ is the set of $p(k,2)=p(k)-p(k-1)$ partitions $P$ of the number $k$ with numbers greater or equal than $2$. Notice that $\mathcal{F(DS)}$ contains single-trace corrections,   corresponding to partitions with exactly one element, and  also multi-trace corrections, corresponding to partitions with more than one element ($|P|\geq 2$). The coefficients $c_1,c_{k,P}$ are arbitrary dimensionless real constants that parameterize and classify the (well defined perturbatively in $\alpha'$) duality and time reparameterization invariant theories $I(\mathcal{S},\Phi,n)$. In particular, the values  $c_1=-\displaystyle\frac{1}{8}$ and  $c_{2,\{2\}}=\displaystyle\frac{1}{64}, \displaystyle\frac{1}{128}$ or $0$ correspond to the  dimensionally reduced low energy effective actions of  the bosonic, heterotic or Type II string theories, respectively, and the higher $c_{k,P}$ are only partially known in string theory.

Since $\mathcal{F(DS)}$ only depends on traces of even powers of $\mathcal{DS}$, it is easy to check that it is a scalar (under time reparameterizations) invariant under global  duality transformations as well as under time-reversal $t\rightarrow -t$.
Therefore the whole action $I(\mathcal{S},\phi,n)$ is $\rO(d,d;\mathbb{R})$, time reparameterization invariant and (up to a sign)  also time-reversal invariant. Thus, the equations of motion are also expected to share these symmetries.

It is convenient to define  a dimensionless scalar function $\widetilde{\mathcal{F}}(\widetilde{\mathcal{X}})\equiv\alpha'\mathcal{F(DS)}$ that only depends on the dimensionless matricial variable $\widetilde{\mathcal{X}}\equiv \sqrt{\alpha'}~\mathcal{DS}$, and verifies $\widetilde{\mathcal{F}}(\widetilde{\mathcal{X}})=\widetilde{\mathcal{F}}(-\widetilde{\mathcal{X}})$.
Then the classical theory described by the action $I(\mathcal{S},\Phi,n)$  can either be studied perturbatively, order by order in $\alpha'$ (assuming infinitesimal values of $\widetilde{\mathcal{X}}^2$), or non-perturbatively.

Finally, notice that  it is possible to add a cosmological constant term $2\Lambda_S=\mathcal{O}(\alpha'^{-1})$   in the definition of $\mathcal{F(DS)}$ when considering the non-perturbative theory. This amounts to adding an $\mathcal{O}(\alpha'^0)$ dimensionless constant $c_0\equiv 2\alpha'\Lambda_S$ to $\widetilde{\mathcal{F}}(\widetilde{\mathcal{X}})$. For instance, $2\Lambda_S=\displaystyle\frac{2(D-D_c)}{3\alpha'}$ in the low energy effective action of non-critical string theory.

\subsection{Equations of motion including multi-trace corrections}\label{sec:eom}

In this section we derive the equations of motion following from \eqref{eqactiontotal}. We generalize the results obtained in \cite{hohm} in order to include the multi-trace corrections, which are necessary to consider solutions with non-vanishing $b$-field or  generic non-isotropic metrics. 

Varying  the action $I(\Phi,\mathcal{S},n)$ with respect to $\Phi,\mathcal{S}$ and $n$,
\begin{equation}\label{eqvar}
\delta_{\Phi,\mathcal{S},n}I=\int dt~n~e^{-\Phi}\left(\delta \Phi E_\Phi+\text{tr}(\delta \mathcal{S}F_\mathcal{S})+\frac{\delta n}{n}E_n\right)\, ,
\end{equation}
one can define  $E_\Phi, F_\mathcal{S}$ and $ E_n$, which are scalars under time reparameterizations.
While $E_\Phi=0$ and $E_n=0$ are the equations of motion for $\Phi$ and $n$,  respectively, $F_\mathcal{S}=0$ is not the equation of motion for $\mathcal{S}$ because the variation $\delta\mathcal{S}$  must verify the conditions $\delta \mathcal{S}=- \mathcal{S}\delta  \mathcal{S} \mathcal{S}$ and $\delta\mathcal{S}^t=\eta\delta \mathcal{S}\eta$, in order to preserve the constraints $\mathcal{S}^2=\mathbf{1}$ and $\mathcal{S}^t=\eta \mathcal{S}\eta$, respectively. 

To impose the constraints on $\mathcal{S}$ and  $\delta S$  we define the projectors
\be\nonumber
\begin{split}
&P_0(A)\equiv \displaystyle\frac{1}{2}(A-\mathcal{S}A\mathcal{S}) ~ {\rm onto ~ the \ subspace \ of \ matrices \ that \  verify }\ A=- {\cal S} A {\cal S}\, ,\\
&P_\pm(A)\equiv \displaystyle\frac{1}{2}(A\pm\eta A^t\eta) ~{\rm onto \ the \  subspace \ of \ matrices \ that \ verify }\  A=\pm\eta A^t\eta\, ,\\
&P_T\equiv P_+\circ P_0=P_0\circ P_+~{\rm onto \ the \ subspace \ of \ matrices \ that \ verify }\ A=- {\cal S} A {\cal S} \ {\rm and }\ A=\eta A^t \eta\, .
\end{split}
\ee
All of these  linear operators $P$ are effectively projectors since $P^2=P$, and they also verify  the property
\be
\tr\left[P(A)B\right]=\tr\left[AP(B)\right]=\tr\left[P(A)P(B)\right]\, , \label{prop}
\ee
for every matrix $A,B \in \mathbb{R}^{2d\times 2d}$. Moreover, considering a local variation $ \delta X_P=P(\delta X)$  constrained to belong to the image of a certain linear projector $P$ (which we assume to verify \eqref{prop}),   the conditions imposed on a matrix $A$ satisfy the following equivalences:
\be \begin{split}
&\displaystyle\int dt~n~\tr(\delta X_P A)=0~\forall~\delta X_P(t)=P(\delta X) \text{ constrained }\\
&\iff \int dt~n~\tr\left[\delta X P(A)\right]=0~\forall~\delta X(t) \text{ unconstrained}\iff P(A)=0 \, . \label{eqprojeom}\end{split}\ee
Thus, the only relevant  information on $A$ for a constrained variation is its projection $P(A)$\footnote{Taking $A=A_1-A_2$ in \eqref{eqprojeom} and using  that $P(A_1-A_2)=P(A_1)-P(A_2)$, one can show that
\be \label{eqfootnote2}
\displaystyle\int dt~n~\tr(\delta X_P A_1)=\int dt~n~\tr(\delta X_P A_2)~\forall~\delta X_P(t) \text{ constrained }\iff P(A_1)=P(A_2)
\ee\label{f2} }. 

In particular, considering $\delta X$ unconstrained, we impose both constraints on $\delta \mathcal{S}$ taking:
\be
\delta \mathcal{S}=P_T(\delta X)=\displaystyle\frac{1}{4}(\delta X+\eta\delta X^t\eta-\mathcal{S}\delta X\mathcal{S}-\mathcal{S}\eta\delta X^t\eta\mathcal{S})\, . \label{deltas}\ee

 Hence, according to \eqref{eqprojeom}, the equation of motion for constrained $\mathcal{S}$ variations  is 
\be
E_\mathcal{S}=P_T(F_\mathcal{S})=0\, , \label{defes}\ee
and using \eqref{prop}, it follows that $\tr(\delta \mathcal{S}F_\mathcal{S})=\tr(\delta XE_\mathcal{S})=\tr(\delta \mathcal{S}E_\mathcal{S})$. This is consistent with the definition $E_\mathcal{S}=P_0(F_\mathcal{S})$ used in \cite{hohm},  because the explicit calculations verify $\eta E_{\mathcal S}^t \eta=E_{\mathcal S}$.

Moreover, noting that  the image of $P_-$ is the $\mathfrak{so}(d,d)=\{\tau \in \mathbb{R}^{2d\times 2d} ~:~\tau\eta+\eta\tau^t=0\}$  Lie algebra, the variation $\delta\mathcal{S}$ in \eqref{deltas} can be written as
\bea
\delta\mathcal{S}=[\tau,\mathcal{S}]\, ,\label{deltatau}
\eea
with $\tau\equiv P_-(\delta X~\mathcal{S}/2)\in \mathfrak{so}(d,d)$. Therefore, we see that every variation $\delta\mathcal{S}$ that preserves the constraints can be written as a local infinitesimal $\rO(d,d;\mathbb{R})$  transformation such as \eqref{deltatau}. Conversely, every local infinitesimal duality transformation of the form $\delta_\tau\mathcal{S}=[\tau,\mathcal{S}]$  preserves the constraints. Indeed, $\delta \mathcal{S}=- \mathcal{S}\delta  \mathcal{S} \mathcal{S}$  is verified  due to  \eqref{deltatau} and $\delta\mathcal{S}=\eta\delta \mathcal{S}^t\eta$ because $\tau\in \mathfrak{so}(d,d)$. 
In particular, for constant $\tau$, $\rO(d,d;\mathbb{R})$ is a global symmetry of the  theory and   there is an associated  conserved Noether charge $\mathcal{Q}$. 

The explicit relation between  $\mathcal{Q}$ and the equation of motion for  $\mathcal{S}$ can be found noting
\be
\delta I=\displaystyle\int dt~n~e^{-\Phi}\tr(\delta \mathcal{S} F_\mathcal{S})=\displaystyle\int dt~n~e^{-\Phi}\tr\left[\tau~2 \mathcal{S} P_0(F_\mathcal{S})\right]\, ,
\ee 
because of \eqref{deltatau}. Then, imposing $\tau \in \mathfrak{so}(d,d)$, one can project $P_-(\mathcal{S}P_0(F_\mathcal{S}))=\mathcal{S}P_+(P_0(F_\mathcal{S}))=\mathcal{S}E_\mathcal{S}$. Further recalling the usual trick to compute the Noether charge, i.e. 
\be\nonumber
\displaystyle\delta_\tau I=\displaystyle\int dt~n~ \text{tr}\left[(\mathcal{D}\tau)\mathcal{Q}\right]=-\displaystyle\int dt~n~ \text{tr}(\tau\mathcal{DQ})\, , \ee
one can take $\mathcal{Q,DQ}\in \mathfrak{so}(d,d)$ since $\tau,\mathcal{D}\tau \in \mathfrak{so}(d,d)$, and use  \eqref{eqfootnote2} in footnote \ref{f2} to find the relation
\be\mathcal{DQ}=-2e^{-\Phi}\mathcal{S}~E_\mathcal{S}\, . \ee
Hence the equation of motion for $\mathcal{S}$  turns out to be equivalent to the conservation of the Noether charge $\mathcal{Q}$. This generalizes the  result found in \cite{hohm} for functions $\mathcal{F(DS)}$ that contain only single-trace corrections, to  generic functions involving  multi-traces. 

 To get the precise expression for $\mathcal{Q}$, it is convenient to define $J$ as
 \be
\delta_\tau I=\displaystyle\int dt~n~ \text{tr}\left[\displaystyle\mathcal{D}(\delta_\tau \mathcal{S})~ J\right]=\int dt~n~\text{tr}\left((\mathcal{D}\tau)[\mathcal{S},J]\right)\, ,
\ee where we used \eqref{deltatau} and
assumed that $J$ is a linear combination of odd powers of $\mathcal{DS}$ (we will show below that this is the case) so that $ [\mathcal{DS},J]=0$. Therefore, using \eqref{eqfootnote2} and $[\mathcal{S},(\mathcal{DS})^{2k-1}]=2\mathcal{S}(\mathcal{DS})^{2k-1} \in \mathfrak{so}(d,d)$, as implied by the constraints on $\mathcal{S}$, we can identify 
\be
\mathcal{Q}=P_-([\mathcal{S},J])=P_-(2\mathcal{S} J)=2\mathcal{S} J\in \mathfrak{so}(d,d)\, .\ee

Varying the explicit form of the action  we get
\begin{equation}    
\begin{split}
\delta_\tau I&=-\displaystyle\int dt~n~e^{-\Phi}~\delta_\tau\left[\mathcal{F(DS)}\right]=-\displaystyle\int dt~n~e^{-\Phi}~\text{tr}\left[\delta_\tau(\mathcal{DS})~\mathcal{F'(DS)}\right]\, ,\\
\end{split}
\end{equation}
where we define the derivative $\mathcal{F}'(A)$ of a scalar function $\mathcal{F}(A)$ with respect to a matrix $A$, as a matrix such that $\delta\left[\mathcal{F}(A)\right]=\tr\left[\delta A~\mathcal{F}'(A)\right]$ \cite{matrixcalculus},
 and then
\begin{equation} J=-e^{-\Phi}~\mathcal{F'(DS)}\implies \mathcal{Q}=-2~e^{-\Phi}~\mathcal{S}~\mathcal{F'(DS)}\, .
\end{equation}
 We see that $J$ is in fact a linear combination of odd powers $\left(\mathcal{DS}\right)^{2k-1}$. Actually, explicitly deriving  the asymptotic expansion \eqref{eqdeffcursiva}, we get
\begin{equation}\label{eqfprimeds}
\begin{split}
\mathcal{F'(DS)}&=-2c_1\mathcal{DS}-\displaystyle\sum_{k=2}^\infty\alpha'\,^{k-1}\displaystyle\sum_{P\in~\text{Part}(k,2)}c_{k,P}\displaystyle\sum_{m_0\in P}2m_0~(\mathcal{DS})^{2m_0-1} \prod_{m\in P-\{m_0\}}\text{tr}\left[(\mathcal{DS})^{2m}\right]\, ,
\end{split}
\end{equation}
where we used $\displaystyle\frac{\partial \left\{\text{tr}\left[g(X)\right]\right\}}{\partial X}=g'(X)$ for $g(X)$ polynomial, since $\delta\left[\tr(X^n)\right]=\tr\left[\delta X~nX^{n-1}\right]$ for all $n\in \mathbb{N}_0$ \cite{matrixcalculus}.
Hence, we confirm that $\mathcal{Q}$ is a linear combination of terms like $\mathcal{S}(\mathcal{DS})^{2m_0-1}\in \mathfrak{so}(d,d)$, and consequently $\mathcal{Q}\in\mathfrak{so}(d,d)$.

We now turn to the simpler equations of motion for $\Phi$ and $n$. The former is trivial
\be
E_\Phi=2\mathcal{D}^2\Phi-(\mathcal{D}\Phi)^2+\mathcal{F(DS)}=0\, .
\ee
To calculate $E_n$, we  consider 
\be
\delta_n\left[\mathcal{F}(\mathcal{DS})\right]=\text{tr}\left[\delta_n(\mathcal{DS})~\mathcal{F}'(\mathcal{DS})\right]=-\displaystyle\frac{\delta n}{n}~\text{tr}\left[\mathcal{DS}~\mathcal{F}'(\mathcal{DS})\right]
\ee
 and 
\be
\delta_n\left[(\mathcal{D}\Phi)^2\right]=-2\displaystyle\frac{\delta n}{n}~(\mathcal{D}\Phi)^2\, ,
\ee which lead to
\be\displaystyle\delta_nI=\int dt~n~e^{-\Phi}~\frac{\delta n}{n}~\left\{(\mathcal{D}\Phi)^2-\mathcal{F(DS)}+\text{tr}\left[\mathcal{DS}~\mathcal{F}'(\mathcal{DS})\right]\right\}\, ,
\ee
from where we can identify $E_n$.

Summarizing, the field equations  including the multi-trace corrections, are
\begin{subequations}\label{eomf}
\begin{align}
&E_n=(\mathcal{D}\Phi)^2-\mathcal{F(DS)}+\text{tr}\left[\mathcal{DS}~\mathcal{F}'(\mathcal{DS})\right]=0\label{eomfn}\\
&E_\Phi+E_n=2\mathcal{D}^2\Phi+\text{tr}\left[\mathcal{DS}~\mathcal{F}'(\mathcal{DS})\right]=0\label{eomfphi}\\
& E_\mathcal{S}=-\frac{1}{2}e^\Phi\mathcal{S}~\mathcal{DQ}=\mathbf{0} \iff \mathcal{DQ}=\mathbf{0}\iff \mathcal{Q}=\text{constant}\in\mathfrak{so}(d,d)
\end{align}
\end{subequations}
Notice that  the first one is a constraint between $\mathcal{D}\Phi$ and $\mathcal{DS}$, while the other two  determine the dynamics of $\Phi$ and $\mathcal{S}$, since they contain second derivatives. All of them  are $\rO(d,d;\mathbb{R})$
and time reparameterization invariant. Furthermore, they are also invariant under time reversal $t\rightarrow -t$ as  expected, since $\mathcal{F(DS)}$ only contains even powers of $\mathcal{DS}$. In other words, for a given solution $\mathcal{S}(t),\Phi(t),n(t)$ there is also a time-reversed solution $\widetilde{\mathcal{S}}(t)\equiv \mathcal{S}(-t),\widetilde{\Phi}(t)\equiv \Phi(-t),\widetilde{n}(t)\equiv n(-t)$.

The Bianchi identity, which follows from the time reparameterization invariance,  is \cite{hohm} 
\begin{equation}\label{eqbianchi}
\mathcal{D}E_n=(\mathcal{D}\Phi) (E_\Phi+E_n)+\tr\left[(\mathcal{DS})E_\mathcal{S}\right]\, 
\end{equation}
Then, if $\mathcal{D}\Phi\neq 0$ for (almost) all times, it is only necessary to solve the equations
\begin{equation}
E_n=0~~,~~E_\mathcal{S}=\mathbf{0}\iff \mathcal{Q}=\text{constant}\, ,
\end{equation}
since they imply (together with the Bianchi identity) that $E_\Phi+E_n=0$.

Moreover, it is always sufficient to consider
\begin{equation}\label{eqnecesbianchi}
E_\Phi+E_n=0~~,~~E_\mathcal{S}=\mathbf{0}\iff \mathcal{Q}=\text{constant}~~,~~E_n(t_0)=0\, ,
\end{equation}
 where $E_n(t_0)$ is evaluated at a certain initial time $t_0$, as they imply  $E_n(t)=0$ for all times.

To solve these equations perturbatively, one should replace $\mathcal{F(DS)}$ and $\mathcal{F'(DS)}$ by their asymptotic expansions up to a certain order. More precisely, one should solve the two-derivative equations and then correct them perturbatively. On the other hand, to find non-perturbative solutions, one should consider $\mathcal{F(DS)}$ as a general scalar function of $\mathcal{DS}$, or more precisely consider $\widetilde{\mathcal{F}}(\widetilde{\mathcal{X}})$ as a general dimensionless scalar function of the dimensionless matrix $\widetilde{\mathcal X}^2$, which may take non-infinitesimal values. 

Notice that adding a cosmological constant $2\Lambda_S$, so that  $\mathcal{F(DS)}\rightarrow \mathcal{F(DS)}+2\Lambda_S$, produces only a constant shift in $\mathcal{F(DS)}$, without changing $\mathcal{F'(DS)}$. Consequently,  the only change in \eqref{eqnecesbianchi} is the initial condition, which becomes $E_n(t_0)-2\Lambda_S=0$.

In the forthcoming sections we will look for dS solutions of these equations, i.e. solutions with a Friedmann-Lemaitre-Robertson-Walker metric with curvature $k=0$ and scale factor $a(t)=e^{H_0t}$, with  constant Hubble parameter $H_0$.

\section{Non-perturbative dS vacua in $n+1\leq d+1$ and $b=\mathbf{0}$} \label{sec:feq}

Setting
 the $b$-field  to zero and the spatial metric to
 $g_{ij}=a(t)^2\delta_{ij}$, it was shown in \cite{hohm} that the equations of motion \eqref{eomf} reduce to the (string) Friedmann equations (as found for instance in \cite{yz}) corrected with higher derivatives. These equations can be integrated  perturbatively to arbitrary order in $\alpha'$ and furthermore, it was argued that they may admit dS solutions  in the string frame  that are non-perturbative in
 $\alpha'$. A necessary condition to have $d$ dimensional dS solutions in the Einstein frame with non-constant dilaton  was found in \cite{krishnan}, in the form of a second order non-linear ordinary differential equation (ODE) to be satisfied by the function that describes the $\alpha'$-corrections. Additionally, $d$ dimensional isotropic dS solutions  were also discussed including  duality covariant matter sources  in \cite{branden}.

In this section we consider the simplest possible extension of the isotropic ansatz,  namely a metric with  one dynamical scale factor $a(t)$ in $n<d$  isotropic spatial dimensions\footnote{It should be clear from the context when $n$ refers to the number of spatial  dimensions or to the $g_{00}=-n^2(t)$ component of the metric.}  and another constant scale factor $a_{0}$  in the other $d-n$ isotropic spatial dimensions, i.e.
\begin{equation}\label{eqansatzfacil}
b=0~~~,~~~a_i=\begin{cases}a(t)~\text{ if  }~~1\leq i\leq n\\
a_{0}={\rm constant}~\text{ if  }n+1\leq i\leq d
\end{cases}
\end{equation}
This ansatz was analyzed in \cite{branden2} in presence of  matter sources.

If one considered different constant scale factors $a_{j,0}$ for each of the  $d-n$ spatial dimensions $x^j$,  a global $\rO(d,d;\mathbb{R})$ transformation  could always be performed,  corresponding to a reparameterization  $x^j\rightarrow x'^j=\displaystyle\frac{a_{0}}{a_{j,0}}~x^j$ \cite{sen91,hsz15}, which amounts to replacing all the  different constant $a_{j,0}$ with a single  one, thus obtaining  \eqref{eqansatzfacil} (the $b$-field is not affected since $b=0$).
The interval  $ds^2=-n^2(t)dt^2+\displaystyle\sum_{i=1}^na^2(t)dx^idx^i+\displaystyle\sum_{j=n+1}^da_{0}^2dx^jdx^j$ certainly has a potential to describe our 4-dimensional universe if $n=3$. In principle, any rotation that mixes the $n$ spatial dimensions having dynamical scale factor  with  the $d-n$ remaining ones is not a symmetry of the theory.

As observed in \cite{branden2}, in this case it is not necessary to include the multi-trace corrections in \eqref{eqdeffcursiva}. The function $\mathcal{F(DS)}$ 
 will be a  single-variable $n$-dependent function $F_n(H)$ of the unique dynamical Hubble parameter $H=\mathcal{D}\ln (a)=\displaystyle\frac{\mathcal{D}a}a$ as in \cite{hohm}.

\subsection{Non-isotropic metric and vanishing \texorpdfstring{$b$}{1}-field}\label{sec:noniso}

In a diagonal metric $g_{ij}=a_i^2(t)\delta_{ij}$  with different  scale factors $a_i(t)$ for each spatial direction $x^i$ and $b=0$, the matrix  $\mathcal{S}$ takes the form
\be
\mathcal{S}=\begin{pmatrix}0&g\\g^{-1}&0\end{pmatrix}=\begin{pmatrix}0&\text{diag}(a_i^2)\\\text{diag}(a_i^{-2})&0\end{pmatrix}\, ,\ee
and its derivative
\be
\mathcal{DS}=2\begin{pmatrix}0&\text{diag}(H_ia_i^2)\\\text{diag}(-H_ia_i^{-2})&0\end{pmatrix}\, \quad {\rm with} \quad
\left(\mathcal{DS}\right)^2=-4\begin{pmatrix}\text{diag}(H_i^2)&0\\ 0&\text{diag}(H_i^{2})\end{pmatrix} \, .  \ee
 $H_i\equiv \mathcal{D}\ln(a_i)$ is the Hubble parameter associated to  $x^i$. Choosing the time parameterization $t_S$ such that $n(t_S)=1$, we can write $H_i=\partial_{t_S}\ln(a_i)$.

In  the simpler ansatz \eqref{eqansatzfacil} with  $g= \diag_{n}(a^2(t),a_0^2)\equiv\diag({\underbrace{a^2(t),\cdots ,a^2(t)}_{n\ \rm times}},{\underbrace{a_0^2,  \cdots,a_0^2}_{(d-n)\ \rm times}})$, the matrix $\mathcal{S}$ simplifies to
\be\mathcal{S}=\begin{pmatrix}0&\diag_n(a^2(t),a_0^2)\\\diag_n(a^{-2}(t),a_0^{-2})&0\end{pmatrix}\, .\label{easyansatz}\ee Its time derivative is
\be
\mathcal{DS}=2\begin{pmatrix}0&\text{diag}_n(Ha^2(t),0)\\\text{diag}_n(-Ha^{-2}(t),0)&0\end{pmatrix}\equiv 2H\mathcal{J}_n\, ,
\ee with $H\equiv \mathcal{D}(\ln(a(t)))$ the only non-trivial Hubble parameter, and
\begin{equation}
\left(\mathcal{DS}\right)^2=-4H^2~\begin{pmatrix}\text{diag}_n(1,0)&0\\ 0&\text{diag}_n(1,0)\end{pmatrix}\equiv -4H^2 \mathcal{I}_n 
\end{equation}
since $\mathcal{J}_n^2=-\mathcal{I}_n$.  Thus, we expect that $\mathcal{F(DS)}$ can be expressed as a single-variable function $F_n(H)$ and  also  that the multi-trace corrections can be absorbed into the single-trace ones.

To prove  this, we compute $\mathcal{F(DS)}$ considering that $\mathcal{I}_n$ is idempotent ($\mathcal{I}_n^2=\mathcal{I}_n\implies \mathcal{I}_n^m=\mathcal{I}_n$). Then $\left(\mathcal{DS}\right)^{2m}=(-1)^m2^{2m}H^{2m} ~\mathcal{I}_n$ and $\text{tr}\left[\left(\mathcal{DS}\right)^{2m}\right]=(-1)^m2^{2m}H^{2m}~2n$, so that
\begin{equation}\label{fds}
\begin{split}
\mathcal{F(DS)}&=8nc_1H^{2}-2n\displaystyle\sum_{k=2}^\infty~\alpha'\,^{k-1}~(-1)^k2^{2k}H^{2k}\displaystyle\sum_{P\in~\text{Part}(k,2)}(2n)^{|P|-1}c_{k,P}\\
&=2n\displaystyle\sum_{k=1}^\infty~\alpha'\,^{k-1}~(-1)^{k-1}2^{2k}H^{2k}c_k^{(n)}\equiv F_n(H)\, ,\\
\end{split}
\end{equation}
where  we absorbed the $c_{k,P}$   in an $n$-dependent  single coefficient  $c_k^{(n)}\equiv\displaystyle\sum_{P\in~\text{Part}(k,2)}(2n)^{|P|-1}c_{k,P}$ for $k\geq 2$ and $c_1^{(n)}\equiv c_1$. This is related to the way in which the multi-trace corrections are absorbed as single-trace corrections  in \cite{hohm}\footnote{More generally,  the multi-trace corrections can be absorbed as single-traces if  $\displaystyle\prod_{m\in P}\tr\left[(\mathcal{DS})^{2m}\right]=d_P\tr\left[(\mathcal{DS})^{2k}\right]$ for every $P\in \text{Part}(k,2)$, with $d_P$ a constant that might depend on $P$. This is always the case if $(\mathcal{DS})^2$ is proportional to an idempotent matrix (with eigenvalues   $0$ and $1$), whose trace must be a constant non-negative integer.}. Indeed, we explicitly expressed  $\mathcal{F(DS)}$ as a single-variable function $F_n(H)$ of the only non-trivial Hubble parameter. Evaluating in $n=d$, we recover the function $F(H)$ defined in  \cite{hohm} (with coefficients $c_k=c_k^{(d)}$).

It will also be useful to compute
\begin{equation}\label{gran}
\begin{split}
&\mathcal{F'(DS)}=-2c_1\mathcal{DS}-\displaystyle\sum_{k=2}^\infty\alpha'\,^{k-1}\displaystyle\sum_{P\in~\text{Part}(k,2)}c_{k,P}\displaystyle\sum_{m_0\in P}(2m_0)~(\mathcal{DS})^{2m_0-1} \prod_{m\in P-\{m_0\}}\text{tr}((\mathcal{DS})^{2m})\\
&=\mathcal{DS}\left[-2c_1\mathbf{1}-\displaystyle\sum_{k=2}^\infty\alpha'\,^{k-1}\displaystyle\sum_{P\in~\text{Part}(k,2)}c_{k,P}\displaystyle\sum_{m_0\in P}(2m_0)~(\mathcal{DS})^{2(m_0-1)} \prod_{m\in P-\{m_0\}}\text{tr}((\mathcal{DS})^{2m})\right]\, .
\end{split}
\end{equation}
Using that $(\mathcal{DS})^{2m}= (-4)^{m}~H^{2m}~\mathcal{I}_n$ and $\tr\left[(\mathcal{DS})^{2m}\right]=2n~(-4)^{m}~H^{2m}$, one can  see that the matrix between brackets in \eqref{gran} is diagonal, with components equal to $-2c_1$ in the elements that correspond to the zeroes of $\mathcal{I}_n$. In addition, a straightforward computation shows that the remaining diagonal components corresponding to the elements $1$ of $\mathcal{I}_n$ are equal to $-\displaystyle\frac{F_n'(H)}{8nH}$. Thus we can express
\begin{equation}
\mathcal{F'(DS)}=-\displaystyle\frac{1}{4n}~F_n'(H)~\begin{pmatrix}0&\text{diag}_n\left(a^2,0\right)\\ \text{diag}_n\left(-a^{-2},0\right)&0\end{pmatrix}= -\displaystyle\frac{1}{4n}~F_n'(H)~\mathcal{J}_n\, .
\end{equation}

The equations of motion can now be calculated, considering that
\be\tr\left[\mathcal{DS~F'(DS)}\right]=-\displaystyle\frac{2H}{4n}~F'_n(H)~\tr(\mathcal{J}_n^2)=\displaystyle\frac{H}{2n}~F'_n(H)~\tr(\mathcal{I}_n)=HF_n'(H)
\ee and
\begin{equation}
\begin{split}
\mathcal{Q}&=-2e^{-\Phi}\mathcal{S}\mathcal{F'(DS)}=\displaystyle\frac{1}{2n}~e^{-\Phi}F_n'(H)~\begin{pmatrix}\text{diag}_n(-1,0)&0\\ 0&\text{diag}_n(1,0)\end{pmatrix}\, .
\end{split}
\end{equation}
Hence $\mathcal{Q}=\text{constant}\in \mathfrak{so}(d,d)\iff q\equiv e^{-\Phi}F_n'(H)=\text{constant}$.
Therefore, in the ansatz \eqref{easyansatz} the equations of motion \eqref{eomf} take the form
\begin{subequations} \label{easytot}
\begin{align}
&E_n=(\mathcal{D}\Phi)^2-F_n(H)+HF_n'(H)=0\label{easyn}\\
&E_\Phi+E_n=2\mathcal{D}^2\Phi+HF_n'(H)=0\label{easyphi}\\
&q\equiv e^{-\Phi}~F_n'(H)=\text{constant}\iff \mathcal{D}(e^{-\Phi}~F_n'(H))=0\label{easyq}
\end{align}
\end{subequations}
These are precisely the $\alpha'$-corrected Friedmann equations  found  in \cite{hohm}  for the isotropic ansatz $g_{ij}=a^2(t)\delta_{ij}$, the only difference being  that the function $F_n(H)$ replaces $F(H)$. Then the perturbative solutions found  for $n=d$  in  \cite{hohm}  also solve the equations for
$n\neq d$, simply replacing $d\rightarrow n$ everywhere (including the $c_{k}^{(n)}$ coefficients, i.e. $c_k=c_k^{(d)}\rightarrow c_k^{(n)}$). In particular, it is easy to see that there are no perturbative dS solutions.

To discuss the non-perturbative solutions, it is convenient to deal with the cases $q\neq 0$ and  $q=0$ separately. Since $q=$constant, these two cases obviously do not overlap, and cover all the possibilities.

The solutions for the case $q\neq 0$ when $n\ne d$ are those found for $n=d$ in section 5.1 of \cite{hohm}, simply replacing $F(H)$ by $F_n(H)$,  or equivalently replacing $d\rightarrow n$ everywhere. In particular,  there is an uninteresting Minkowski solution with $H=0=$ constant, but  no dS cosmologies in the string frame.

Instead, the more interesting case $q=0$ that we will explore  in the forthcoming sections,  turns out to  contain many dS solutions.

\subsection{dS solutions in the string frame  } 

If $q= 0$, the equation $e^{-\Phi} F_n'(H)=q$ necessarily implies that
\begin{equation}
F_n'(H)= 0 ~\text{for all times}~\implies H=H_0=\text{constant}
\end{equation}
i.e. $H(t)$ is a constant $H_0$ that is a zero of $F_n'(H)$. If $H(t)$ were not a constant, then the equation $F_n'(H(t))=0$ would be valid in an open neighborhood of a certain $H(t_0)=H_0$. In this case, $F_n'(H)$ must be the zero function $F_n'(H)=0$ for all $H$, which is absurd since we are considering that the asymptotic expansion of $F_n'$ is non-trivial.

The fact that $F_n'(H)=0$ implies that the conservation of $q=e^{-\Phi}F_n'(H)=0$ is trivial for any function $\Phi(t)$. Then no more information can be obtained from the equation $E_\mathcal{S}=0$ and we turn to the other equations of motion, namely
\begin{equation}
0=E_\Phi+E_n=2\mathcal{D}^2\Phi+HF_n'(H)=2\mathcal{D}^2\Phi \implies \mathcal{D}^2\Phi=0 \iff \mathcal{D}\Phi=-c=\text{constant} \in \mathbb{R} \, ,
\end{equation}
where we used $F_n'(H)=0$ and  defined the real constant $c$ (with $\sign(c)=-\sign(\mathcal{D}\Phi)=\pm$), and
\begin{equation}
0=E_n=(\mathcal{D}\Phi)^2-F_n(H)+HF_n'(H)=c^2-F_n(H) \implies F_n(H)=c^2=\text{constant}\geq 0 \, .
\end{equation}
We conclude that the solutions with $q=0$ are those with $H=H_0=\text{constant}$ such that
\begin{equation}\label{eqsimpleds}
F_n'(H_0)=0~~,~~F_n(H_0)=c^2\geq 0~~,~~\mathcal{D}\Phi=-c~~,
\end{equation}
for some constant $c$ that may take any real value.

Since $H=H_0=\text{constant}$, these are all dS solutions in the string frame. They are non-perturbative because there are non-trivial $\alpha'$-corrections that must mix with each other in order to ensure that $F_n'(H_0)=0$ and $F_n(H_0)=c^2\geq 0$. They are described by the dimensionful constants $(H_0,c)$ that may be measured (non-perturbatively) in units of $1/\sqrt{\alpha'}$.

Notice that the time-reversal of one of these solutions $(H_0,c)$ is a new dS solution with $(-H_0,-c)$, that trivially verifies \eqref{eqsimpleds} since $F_n(-H_0)=F_n(H_0)$ and $F_n'(-H_0)=-F_n'(H_0)$. Moreover, inverting the scale factor $a(t)\leftrightarrow a(t)^{-1}$ of a dS solution $(H_0,c)$, which is a symmetry included in $\rO(d,d)$, another dS solution described by $(-H_0,c)$ is obtained. Hence, from one of these dS solutions, one can always construct another one that is expanding in the string frame, by choosing $H_0>0$, and that also verifies $c>0$ for example (considering $H_0,c\neq 0$).

To the best of our knowledge, these $q=0$ solutions with $c\neq 0$ have not been considered previously in the literature. A detailed analysis of  this case is presented in the next sections, where we will find an interesting zoo of stable and unstable dS geometries that can be dS also in the Einstein frame.

\subsection{dS solutions in the Einstein frame }\label{subsec:eins}

In the case $q=0$, non-perturbative dS solutions with $H=$constant $\neq 0$ and $\mathcal{D} \Phi=0$ were obtained in \cite{hohm} for the isotropic ansatz (i.e. $n=d$).
These are dS metrics in the string frame, which can be trivially generalized to the case $n<d$ simply replacing $F(H)$ by $F_n(H)$. However, they lead to a time dependent Hubble parameter in the Einstein frame and hence do not correspond to the  dS geometries that describe the observable universe. 

To see this, recall the standard  Weyl rescaling of  the  metric that relates the string and  Einstein frames
\be
G_{\mu\nu}=e^{-4\phi/(d-1)}g_{\mu\nu} \, .\label{wr}
\ee
For metrics of the form  \eqref{eqgbphi}, we have $-n_E^2\equiv G_{00}=-e^{-4\phi/(d-1)}n^2$. Thus
 an Einstein frame time covariant derivative can be defined as $\mathcal{D}_E\equiv \displaystyle\frac{1}{n_E(t)}\displaystyle\frac{\partial}{\partial t}$, with the same properties as $\mathcal{D}$, since $n_E(t)$ is trivially a density under time reparameterizations. Both time covariant derivatives are related as $\mathcal{D}_E=e^{2\phi/(d-1)}\mathcal{D}$. We  can always choose a time parameterization $t_E$ such that $n_E(t_E)=1 \iff \mathcal{D}_E= \displaystyle\frac{\partial}{\partial t_E}$, thus $n(t_E)=e^{2\phi(t_E)/(d-1)}\iff \mathcal{D}=e^{-2\phi(t_E)/(d-1)}\displaystyle\frac{\partial}{\partial t_E}$.

Restricting to diagonal metrics $g=\diag(a_i^2(t))$, the scale factors are related as
\be a_{E,i}(t)=e^{-2\phi/(d-1)}a_i(t)\, .
\ee
Then the Hubble parameter  associated to the $x^i$ direction in the Einstein frame is
\be \label{eqhubbleeinsteinfr}
\begin{split}
H_{E,i}&\equiv \mathcal{D}_E(\ln(a_{E,i}(t)))=e^{2\phi/(d-1)}\left(H_i-\displaystyle\frac{2~\mathcal{D}\phi}{d-1}\right)\, .
\end{split}
\ee In particular, for a constant dilaton $ \phi=\phi_0=$constant, the Weyl rescaling \eqref{wr} just amounts to multiplying the metric by a global constant. Thus, a dS metric in the string frame  with Hubble parameters $H_i=H_{i,0}=$constant  is also a dS metric in the Einstein frame  with Hubble parameters $H_{E,i}=H_{E,i,0}=e^{2\phi_0/(d-1)}H_{i,0}=$constant. 

Therefore,  the dilaton $\phi$ cannot be constant  in a solution with
 $H_0=$ constant $\neq 0$ and $\mathcal{D}\Phi=0$  in the string frame, since $ \mathcal{D}\phi= \mathcal{D}\Phi+nH_0\ne0$. Hence this solution
leads to  a time dependent Hubble parameter in the Einstein frame, and there is no proper dS geometry in this case. 

Instead, the solutions $(H_0,c)$  with $\mathcal{D}\Phi=-c=$ constant described by \eqref{eqsimpleds}
 admit dS cosmologies with constant dilaton when $c\ne 0$. Indeed,
imposing the condition
\begin{equation}
2\mathcal{D}\phi=\mathcal{D}\Phi+\mathcal{D}(\ln(\sqrt{\det{g_{ij}}}))=\mathcal{D}\Phi+\displaystyle\sum_{i=1}^d H_i=0\, 
\end{equation}
 in the ansatz $g=\diag_n(a^2(t),a_0^2)$ and $b=0$, amounts to
\begin{equation}\label{eqdsconstantphi}
0=2\mathcal{D}\phi=\mathcal{D}\Phi+nH_0+(d-n)\cdot 0=-c+nH_0\iff c=nH_0 \, .
\end{equation}
Thus,  the non-perturbative dS solution $(H_0,c)=(H_0,nH_0)$ in the string frame has constant dilaton $\phi=\phi_0$.
In the Einstein frame, this corresponds to a dS geometry with constant $H_E=H_{E,0}=e^{2\phi_0/(d-1)}H_0$ in  $n$ spatial dimensions and null Hubble parameter in the remaining $d-n$ spatial dimensions. If this was a solution of  string theory, the  string coupling could be taken $g_s=e^{\phi_0}=$constant$\ll1$  for all times, consistently with  string perturbation theory at  genus zero. 

This result seems to contradict the no-go theorem  of \cite{hohm}, which states that  there are no dS solutions in the
Einstein frame with constant dilaton $\phi$. However,  the new solutions \eqref{eqdsconstantphi} have  $\Phi\neq$ constant (i.e. $c\neq 0$), which violates the hypothesis of the theorem, namely
 that the only dS solutions in the string frame are those with $\Phi=$constant (i.e. $c=0$). Then we see that the case $q=0$ is a source of dS solutions in both frames (i.e. $\phi=$ constant $\iff c=nH_0$, in particular $c=nH_0\neq 0$). 

Notice that for every dS solution $(H_0,c)$ with constant dilaton (i.e. $c=nH_0$), there is its time-reversed dS solution $(-H_0,-c)$ with constant dilaton since $-c=n(-H_0)$. Thus, there are two types of dS solutions with constant dilaton: the expanding cosmologies $(H_0,c=nH_0)$ with $H_0>0$ and the contracting ones $(-H_0,-c=n(-H_0))$ with $-H_0<0$.

On the other hand,  a dS solution in the string frame described by $(H_0,c)$ with non-constant dilaton must have either $c>nH_0 \iff \mathcal{D}\phi=\text{constant}<0$ or $c<nH_0\iff \mathcal{D}\phi=\text{constant}>0$. In the former (latter) case, the string coupling would be small only for late (early) times. The Einstein frame Hubble parameter of these solutions is of the form
\be \label{se} \begin{split}
H_E&=\frac{e^{2\phi(t)/(d-1)}}{d-1}\left[c+(d-n-1)H_0\right] \, ,
\end{split}
\ee
with $\displaystyle\phi(t)=\phi_0+(nH_0-c)\int_{t_0}^t dt~n(t)$. Then $H_E$ is a non-zero constant  if and only if $\phi=$constant $\iff c=nH_0$. If $c\ne nH_0$, an expanding (contracting) solution in the Einstein frame (which is not dS) is obtained  only if $c+(d-n-1)H_0$ is positive (negative), see Figure \ref{ds_str_eins}.

Alternatively, one could search for  dS geometries in the Einstein frame with non-constant dilaton 
(i.e. that do not verify \eqref{eqsimpleds} and are not dS in the string frame).  In general, they have to satisfy quite non-trivial conditions.
In the case of  isotropic metric  $g_{ij}=a^2(t)\delta_{ij}$ and $b=0$, the function $F(H)=F_{n=d}(H)$ that describes the $\alpha'$-corrections has to verify
the non-linear second order ODE 
\begin{equation}\label{ode}
\begin{split}
&-\displaystyle\frac{(d+1)}{2}~HF'(H)+F(H)+dH^2=\pm \sqrt{-HF'(H)+F(H)}\left((d-1)\frac{F'(H)}{F''(H)}+(d+1)H\right)\\
\end{split}
\end{equation}
with $\pm=-\sign(\mathcal{D}\Phi)$ \cite{krishnan}. This must be regarded as an ODE because $H=H(t)\neq $ constant can take any value in an open neighborhood of a certain $H(t_0)$, provided the dilaton $\phi$ is not constant. A possible solution  is  $F(H)=-H^2+BH$ for all $B\in\mathbb{R}$ if $\pm=\sign(H)$, but this does not correspond to string theory. Even in the case $B=0$ (which ensures $F(H)=F(-H)$), the $\mathcal{O}(H^2)$ term does not agree with $F(H)=-dH^2+\cdots$ in the asymptotic expansion \eqref{fds}. Besides this one, there should be another kind of solution with another integration constant (rather than just $B$), which should be an \textit{acceptable} function $F(H)$ according to \cite{krishnan} in order to be a possible non-perturbative dS solution of tree level string theory. For instance, the non-constant string coupling  must verify $g_s=e^{\phi}\ll 1$ at all times.

Notice that  dS geometries in the Einstein frame with constant dilaton correspond to $H=H_0=$ constant $\neq0$ in the string frame, in particular in the case $n=d$. Thus, the only possible solutions for isotropic metric are those with $q=0$, i.e. $F'(H_0)=0$, $F(H_0)=c^2$ and $\mathcal{D}\Phi=-c$.  Replacing this in the ODE \eqref{ode}, with $\pm=-\sign(\mathcal{D}\Phi)=\sign(c)$, leads to
\be
c^2+dH_0^2=\sign(c)|c|(d+1)H_0=(d+1)cH_0 \iff 0=c^2-(d+1)H_0c+dH_0^2=(c-dH_0)(c-H_0)\, .\nonumber\ee Then, there are two possible solutions. Either $c=H_0$, which implies 
\be \nonumber
(d-1)H_0-2\mathcal{D}\phi=(d-1)H_0-(-c+dH_0)=c-H_0=0\implies H_E=0\, ,\ee i.e. a Minkowski metric in the Einstein frame; or $c=dH_0$, which is the dS solution with constant dilaton   \eqref{eqdsconstantphi} (in the case $n=d$). But now we found it by a different pathway, i.e. solving an algebraic equation instead of a differential one. Indeed, for both  constant $H=H_0$ and $\phi$,   the ODE becomes  algebraic.

\subsection{Stability and types of dS solutions}\label{subsec:stability} 

Following \cite{stabilityhf}, to study the stability of the dS solutions found above, it is convenient to define $y\equiv\mathcal{D}\Phi$ and  recall  that the equations of motion   \eqref{easyq} and \eqref{easyphi}  are first order differential equations for $y$ and $H$, while  \eqref{easyn}  is a constraint between them.
Therefore, a variation of the dynamical variables $\delta y=\delta(\mathcal{D}\Phi)$ and $\delta H$  must preserve the constraint \eqref{easyn}, 
\be \label{eqstabconstrvar}
0=\delta E_n=2y~\delta y+HF_n''(H)~\delta H\, .
\ee
Thus, the variation of the first derivatives $\mathcal{D}y$ and $\mathcal{D}H$ under  $\delta y$ and $\delta H$ are determined by the equations \eqref{easyphi} and \eqref{easyq} as
\be
\begin{split}
\delta (\mathcal{D}y)&=-\displaystyle\frac{1}{2}~(F_n'(H)+HF_n''(H))~\delta H\, ,\\
F_n''(H)\delta (\mathcal{D}H) &=F_n'(H)~\delta y+yF_n''(H)~\delta H-F_n'''(H)\mathcal{D}H~\delta H\, .
\end{split}
\ee
Evaluating for the solutions  \eqref{eqsimpleds} described by $(H_0,c)$ in the string frame and imposing $\mathcal{D}H=\mathcal{D}(H_0)=0$, $F_n'(H_0)=0$ and $y=\mathcal{D}\Phi=-c$, we can rewrite the variations as
\be
\delta (\mathcal{D}y)=-c~\delta y\, \qquad {\rm and}\qquad
\delta (\mathcal{D}H)=-c~\delta H\, ,
\ee
where we also applied the constraint \eqref{eqstabconstrvar}, and assumed $F_n''(H_0)\neq 0$.

Therefore, we see that a dS solution in the string frame described by $(H_0,c)$ is stable if $c>0$, and  unstable if $c<0$, for both the dynamics of $y$ and $H$. 
Notice that if $H_0,c\neq 0$ and if $\delta y$ can take non-zero values, the constraint \eqref{eqstabconstrvar} necessarily implies $F_n''(H_0)\neq 0$. In the case $c=0$ (in which $F_n''(H_0)= 0$ if $\delta H$ can take non-zero values, because of \eqref{eqstabconstrvar}), further analysis is required to establish the stability or not of the solution.

\begin{figure}[H]
    \centering
    \subfloat[dS solutions $(H_0,c)$ in the string frame] {\includegraphics[width=0.49\textwidth]{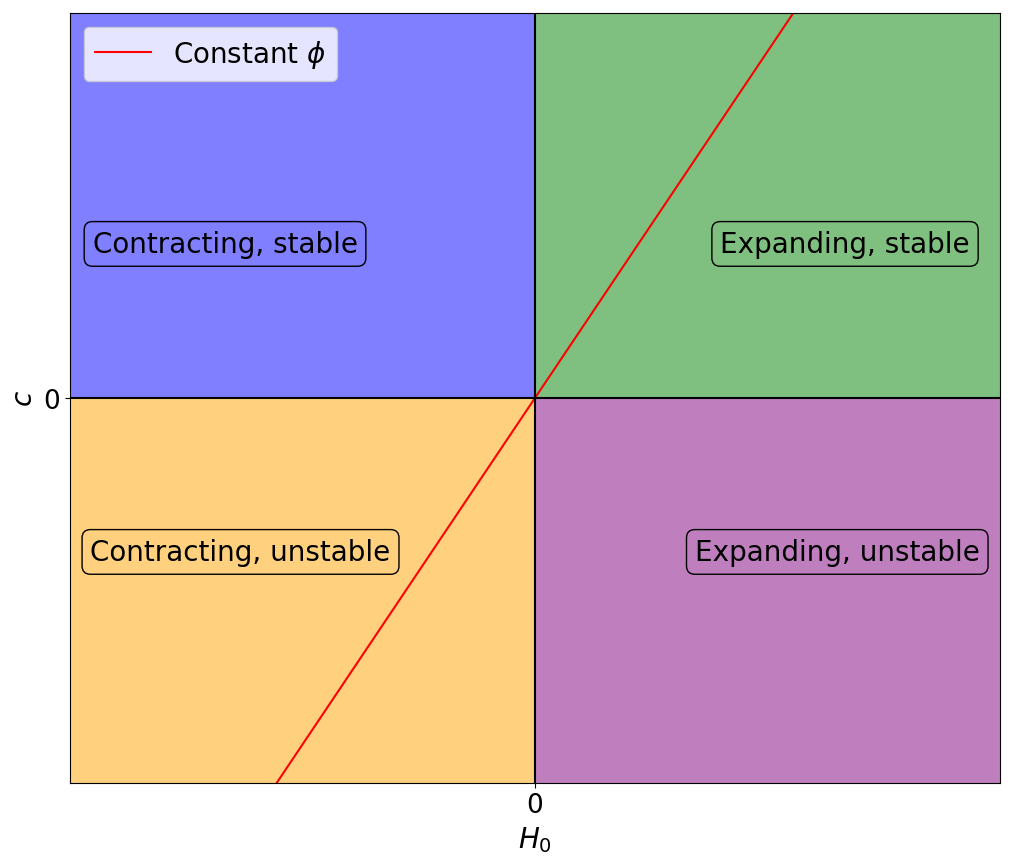}\label{ds_str}}
    \hfill
    \subfloat[dS solutions $(H_0,c)$ as seen in the Einstein frame]{\includegraphics[width=0.49\textwidth]{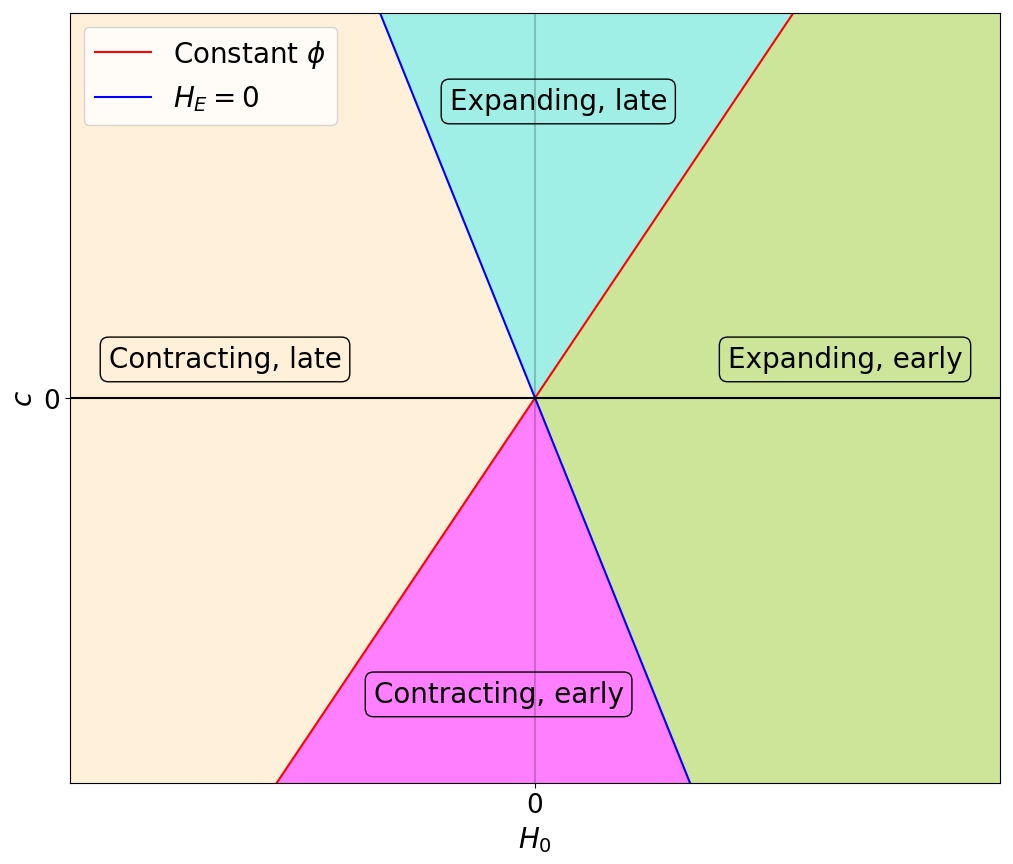}\label{ds_str_eins}}
    \caption{ $(a)$ 
expansion $H_0>0$/contraction $H_0<0$ and stability.
The red line is the region $\phi=$ constant $\iff c=nH_0$. $(b)$ expansion $H_E(t)>0$/contraction $H_E(t)<0$. Late and early refer to the times for which $g_s=e^{\phi(t)}\ll 1$. The blue line is the region $H_E=0\iff c=-(d-n-1)H_0$ and the red line is the region $\phi=$ constant with $g_s=e^{\phi_0}\ll 1$.}
    \label{ds_figs}
\end{figure}

Then, there are four types of dS solutions in the string frame with $H_0,c\neq 0$ (see Figure \ref{ds_str}): An expanding stable dS solution $(H_0,c)$ with $H_0>0,~c>0$; its time-reversed solution $(-H_0,-c)$ which is contracting and unstable; the solution $(-H_0,c)$ that is obtained from a scale factor inversion which is contracting and stable; and the solution $(H_0,-c)$ that is obtained from a scale factor inversion and a time-reversion which is expanding and unstable. Hence,  an expanding stable dS solution in the string frame  can always be chosen.

In particular, a dS solution $(H_0,c=nH_0)$ with constant dilaton is stable if $c=nH_0>0$, i.e. if it is an expanding solution. Indeed, the condition $c=nH_0$ implies $\sign(H_0)=\sign(c)$. Thus, every non-perturbative expanding dS solution with constant dilaton is stable.

We summarize the properties of the dS solutions $(H_0,c)$ in the string and Einstein frames (expansion/contraction, stability, and times for which $g_s=e^{\phi(t)}\ll 1$)  in Figure \ref{ds_figs}. Note that for $n<d$, every stable expanding  dS solution in the string frame (green quadrant of  Figure \ref{ds_str}), corresponds to a stable expanding geometry $H_E(t)>0$  in the Einstein frame (cyan and light green regions in Figure \ref{ds_str_eins}), while for $n=d$ this only happens if $c>H_0$. Recall the precise relation between the Hubble parameters given in \eqref{se}.

\section{Generalized ansatz of commuting matrices}\label{sec:matrixds2}

Having solved the equations of motion in a simplified setting that  involves a diagonal metric and vanishing $b$-field, we now extend the analysis for more general  fields.  We first discuss a generalized ansatz in which the matrices $g,\mathcal{D}(g),\mathcal{D}b$ are taken to commute among each other, and show that with this assumption, the $\rO(d,d)$ symmetry allows  to take a diagonal metric and  block diagonal matrices $b,\mathcal{D}b$, without loss of generality. 
We then show that  in this case the equations of motion can be diagonalized, which greatly simplifies the search of solutions.

In order to examine the $\alpha'$-complete cosmology  in a more general setting, since the functions $\mathcal{F(DS)}$ and $\mathcal{F'(DS)}$ only involve
 traces of even powers of $\mathcal{DS}$, it is convenient to first  compute the matrix $(\mathcal{DS})^2$. 
This is a straightforward calculation,  and after some tedious 
manipulations, the result can always be written as
\begin{equation}\label{eqdssqgeneral}
\begin{split}
\left(\mathcal{DS}\right)^2=\begin{pmatrix}A&bg^{-1}Ag-Ab+(g-bg^{-1}b)Cg\\ C&g^{-1}Ag-g^{-1}bCg-Cb\end{pmatrix}\, ,
\end{split}
\end{equation}
where  $A\equiv \mathcal{D}(g)\mathcal{D}(g^{-1})+(\mathcal{D}(b)g^{-1})^2+bC$ and $C\equiv\mathcal{D}(g^{-1})\mathcal{D}(b)g^{-1}-g^{-1}\mathcal{D}(b)\mathcal{D}(g^{-1}) $   contain all the time derivatives.

This expression seems quite complicated. 
However, if the matrices $g,\mathcal{D}(g),\mathcal{D}b$ are taken to commute among each other, then $g^{-1},\mathcal{D}(g^{-1})$ also belong to such set of commuting matrices, and 
 $C=\mathbf{0}$ while $A=-g^{-2}(\mathcal{D}g)^2+g^{-2}(\mathcal{D}b)^2$ is a symmetric  negative semi-definite matrix.  In this case,  \eqref{eqdssqgeneral} simplifies to
\be\label{eqdssqgenansatz}
\left(\mathcal{DS}\right)^2=\begin{pmatrix}A&[b,A]\\ 0&A\end{pmatrix} \, .
\ee

The assumption that the matrices  commute is a somewhat general ansatz. Nevertheless,  exploiting the $\rO(d,d)$ global invariance (in particular the invariance under global rotations of the spatial coordinates, and constant shifts in the $b$-field \cite{sen91,hsz15}, which preserve the assumption of the general ansatz), we will see that in this case the metric $g$ can  be always made diagonal, and  $\mathcal{D}b$ can be taken to a block diagonal form, without loss of generality.

Indeed, since $g$ and $\mathcal{D}g$ are real symmetric matrices and commute, there is an orthogonal matrix $R=R(t)$, that may depend on time, such that $g=Rg_DR^{-1}$ and $\mathcal{D}g=RD_g R^{-1}$, with diagonal matrices $g_D$ and $D_g$  for all times. Taking the time derivative of the first equality,
\be
\mathcal{D}g=\mathcal{D}(Rg_DR^{-1})=R\left(\mathcal{D}g_D+[R^{-1}\mathcal{D}R,g_D]\right)R^{-1}\, ,
\ee
where we used $\mathcal{D}(R^{-1}R)=0$, and using the second one, we get
\be \nonumber
\mathcal{D}g_D+[R^{-1}\mathcal{D}R,g_D]=D_g \, .
\ee
Since $\mathcal{D}g_D$ and $D_g$ are diagonal matrices, $[R^{-1}\mathcal{D}R,g_D]$ is  also  diagonal. E.g., for matrix elements $(g_D)_{ij}=a_i^2(t)\delta_{ij}$, the commutator is
 $[R^{-1}\mathcal{D}R,g_D]_{ij}=(R^{-1}\mathcal{D}R)_{ij}(a^2_j-a^2_i)$. In particular, the diagonal elements  $[R^{-1}\mathcal{D}R,g_D]_{ii}$ are zero, and since $[R^{-1}\mathcal{D}R,g_D]$ is  diagonal, we get
\be
[R^{-1}\mathcal{D}R,g_D]=\mathbf{0} \iff \mathcal{D}g_D=D_g=\diag(2a_i^2 H_i) \, .
\ee
Moreover,  if $a_i\neq a_j$,  then $[R^{-1}\mathcal{D}R,g_D]=\mathbf{0} \implies (R^{-1}\mathcal{D}R)_{ij}=0$.

Let us take, for instance,  the diagonal metric 
\be\label{block}
g_D={\rm diag}(a_{1}^2(t) \mathbf{1}_{s_1},a_2^2(t)\mathbf{1}_{s_2},\cdots, a_{r_b}^2(t)\mathbf{1}_{s_{r_b}})\, ,\ee
where each one of the $r_b$ blocks $a_r^2(t)~\mathbf{1}_{s_r}$ is proportional to the $s_r\times s_r$ identity matrix $\mathbf{1}_{s_r}$ and  $ a_r(t)\neq a_{r'}(t)$ if $r\neq r'$. In other words, all the directions with the same scale factor for all times are grouped in the same block.
From $(R^{-1}\mathcal{D}R)_{ij}=0$ if $a_i\neq a_j$, it follows that
\be\label{difeq}
-R^{-1}\mathcal{D}R=\mathcal{D}(R^{-1})R\equiv E(t) ={\rm diag}(E^{(1)}(t),
   E^{(2)}(t),\cdots, E^{(r_b)}(t)) \, ,
\ee
i.e., $E(t)$ is a block diagonal matrix, with blocks $E^{(r)}(t)$ that correspond to the identity blocks $g^{(r)}=(a^{(r)})^2~\mathbf{1}_{s_r}$ of \eqref{block}. We will refer to these blocks as blocks of equal scale factor $a^{(r)}$.

The   initial condition for $R(t)$ can always be taken to the form $R(t_0)=R^{-1}(t_0)=$ Id with a global
$\rO(d,d)$ transformation, corresponding to a rotation with a constant orthogonal matrix $R_0=R(t_0)^{-1}$, that transforms the fields as $g\rightarrow R_0gR_0^{-1}$ (i.e. $R(t)\rightarrow R_0R(t)$), $b\rightarrow R_0bR_0^{-1}$, so that $R(t_0)=$ Id. Hence, the differential equation  \eqref{difeq} is the  Schrodinger equation for the time evolution operator (with $U(t)= R^{-1}(t)$ and the Hamiltonian  $(i\hbar)^{-1}H(t)= n(t)E(t)$), with the initial condition $R^{-1}(t_0)=$Id. Then, it can be formally solved  with a Dyson series:
\be
R^{-1}(t)=\mathcal{T}\left\{\exp\left[\int_{t_0}^t dt~n(t)~E(t)\right]\right\} \, ,
\ee
where $\mathcal{T}$ is the time-ordering operator.  Consequently, the orthogonal matrix $R^{-1}(t)$ (as well as $R(t)$)  must also be a block diagonal matrix with the same blocks as $E(t)$ and $g_D$.

Therefore, considering that each block of $g_D$ is proportional to the identity and that $R(t),R^{-1}(t)$ are block diagonal, the metric $g=g_D$  can be taken to be  \eqref{block}, which we will do from now on.

Regarding  the $b$-field, since $\mathcal{D}b$ commutes with $g=g_D$,  it must be a block diagonal matrix with blocks  that correspond to the identity blocks \eqref{block} of $g$ (i.e. blocks of equal scale factor):
\be
\mathcal{D}b={\rm diag}(\mathcal{D}b^{(1)},\mathcal{D}b^{(2)},\cdots,\mathcal{D}b^{(r_b)})\, ,\ee
with each block $\mathcal{D}b^{(r)}$ real and antisymmetric. With a global $\rO(d,d)$  transformation, corresponding to a constant shift in the $b$-field,  we can always take $b(t_0)=\mathbf{0}$, without loss of generality, and $b(t)=\displaystyle\int_{t_0}^t dt'~n(t')~\mathcal{D}b(t')$ must also be a block diagonal matrix with  blocks of equal scale factor $a^{(r)}$. Therefore, $b(t)$ commutes with $g,g^{-1},\mathcal{D}g$, and also $[b,A]=g^{-2}[b,(\mathcal{D}b)^2]$.

Summarizing, the generalized ansatz of commuting matrices allows to take,  without loss of generality,  a diagonal metric $g$ and a block diagonal  $\mathcal{D}b$  (with  blocks of equal $a^{(r)}$) with $b(t_0)=\mathbf{0}$, which implies that $b(t)$ commutes with $g,g^{-1},\mathcal{D}g$. 
This includes several interesting cases, such as a generic diagonal metric $g_{ij}=a_i^2(t)~\delta_{ij}$ and  $b=\mathbf{0}$;  an isotropic metric $g_{ij}=a^2(t)~\delta_{ij}$ and  generic $b\ne\mathbf{0}$; a metric with two (or more) dynamical scale factors, e.g. $g=\diag_n(a^2_1(t),a^2_2(t))$  and a  block diagonal $\mathcal{D}b\ne\mathbf{0}$ with the same two (or more) blocks as $g$; etc.

\subsection{Equations of motion}\label{sec:eom2}

The equations of motion that follow from the ansatz of commuting matrices are worked out in the appendix \ref{app:geneom}. They take the form
\begin{subequations}\label{feoms}
\begin{align}
&E_n=(\mathcal{D}\Phi)^2-\mathcal{F}_a(A)+2\tr(A\mathcal{F}_a'(A))=0\, ,\\ 
&E_\Phi+E_n=2\mathcal{D}^2\Phi+2\tr(A\mathcal{F}_a'(A))=0\, ,\\
&{\mathcal{Q}}_2\equiv e^{-\Phi}~g^{-1}\mathcal{D}(g)\mathcal{F}_a'(A)-b{\mathcal{Q}}_1=\text{constant}\iff\mathcal{D}\left(e^{-\Phi}~g^{-1}\mathcal{D}(g)\mathcal{F}_a'(A)\right)-\mathcal{D}b~{\mathcal{Q}}_1=\mathbf{0}\, ,\label{eqc1-4s}\\
&{\mathcal{Q}}_1\equiv e^{-\Phi}~g^{-2}(\mathcal{D}b)\mathcal{F}_a'(A)=\text{constant} \iff \mathcal{D}\left(e^{-\Phi}~g^{-2}(\mathcal{D}b)\mathcal{F}_a'(A)\right)=\mathbf{0}\, ,\label{eqa3-4s}
\end{align}
\end{subequations}
where ${ \mathcal F}({\mathcal {DS}})\equiv {\mathcal F}_a(A)$ since it only depends on $\tr(\mathcal{DS}^{2m})=2\tr(A^m)$, and ${\mathcal{Q}}_2$ and ${\mathcal{Q}}_1$ are  constant symmetric and antisymmetric  integration matrices. The last two equations are the equations of motion for $\mathcal S$ variations,  equivalent to the conservation of the Noether charge
\be\label{eqqgeneralizedansatzs}
\displaystyle{\mathcal{Q}}\equiv -2\begin{pmatrix}{-\mathcal{Q}}_2&-g^2(t_0){\mathcal{Q}}_1
\\ {\mathcal{Q}}_1&{\mathcal{Q}}_2\end{pmatrix}\, .
\ee

As shown in \ref{subapp:diag}, the equations \eqref{feoms} may be diagonalized in terms of a complex unitary matrix $U(t)$, and they become
\begin{subequations}\label{eqsmotion-6s}
\begin{align}
E_n&=(\mathcal{D}\Phi)^2-\mathcal{F}_a(-4D^2)+2\tr((-4D^2)~\mathcal{F}_a'(-4D^2))=0\, ,\label{eqsmotion-6sn}\\ 
E_\Phi+E_n&=2\mathcal{D}^2\Phi+2\tr((-4D^2)~\mathcal{F}_a'(-4D^2))=0\, ,\label{eqsmotion-6snphi}\\
D_{1}&=e^{-\Phi}~g^{-2}(D_b)\mathcal{F}_a'(-4D^2)=\text{ constant}\, ,\label{eqa3-6s}\\
\mathbf{0}&=D_b~D_{1}+\mathcal{D}\left(e^{-\Phi}~g^{-1}\mathcal{D}(g)\mathcal{F}_a'(-4D^2)\right)\, ,\label{eqc1-6s}\\
\mathbf{0}&=[U^{-1}\mathcal{D}U,D_{1}]=\mathcal{D}(g)~[U^{-1}\mathcal{D}U,\mathcal{F}_a'(-4D^2)]\, ,\label{equnitaryus}
\end{align}
\end{subequations}
where  $D_b$, $D_{1}$ and $D^2$ are the real diagonal matrices (the latter with non-negative elements):
\bea
\mathcal{D}b(t)=U(t)(iD_b(t))U^{-1}(t)\, ,\qquad
{\mathcal{Q}}_1=U(t)(iD_{1})U^{-1}(t)\, , \ \ \  \label{eqdbdiag}\\
 \ \ \  A=U(t)(-g^{-2}(\mathcal{D}g)^2-g^{-2}D_b^2)U^{-1}(t)\equiv U(t)(-4D^2)U^{-1}(t)\, ,
\eea

 Note that  the  equations \eqref{eqsmotion-6sn}-\eqref{eqc1-6s}, which determine $\Phi$, the metric and $D_b$,  can be solved without any regard of the unitary matrix $U(t)$. Then, it is  convenient to
 separate them from \eqref{equnitaryus}, which determines $U(t)$, and hence the basis in which $\mathcal{D}b$ is written.
In particular, we can take $U(t)=U_0=$ constant as a valid solution of such equation.

Using  $\rO(d,d)$ invariance, we can replace $\mathcal{D}b(t)\rightarrow R_{b,0}\mathcal{D}b(t)R_{b,0}^{-1}$, or equivalently  $U(t)\rightarrow R_{b,0}U(t)$, where $R_{b,0}$ can be any block diagonal constant orthogonal matrix, with blocks of equal scale factor (in order to preserve the form of $g,\mathcal{D}g,g^{-1}$). Taking $U(t)=U_0=$constant,  $R_{b,0}$ can be chosen such that each block $\mathcal{D}b^{(r)}$ of $\mathcal{D}b$ is a block diagonal matrix
\be\label{eqdb-antis-blockr}
\mathcal{D}b^{(r)}(t)={\rm diag}\left(\beta_{1}^{(r)}~i\sigma_2,\beta_{3}^{(r)}~i\sigma_2,\cdots,\beta_{2c_r-1}^{(r)}~i\sigma_2, \mathbf{0}\right)\, ,
\ee
 with $c_r$ $2\times 2$ blocks   proportional to the Pauli matrix $\sigma_2$,  or a zero  $(s_r-2c_r)\times(s_r-2c_r)$  block.
The eigenvalues of $-i\mathcal{D}b$ are of the form $\pm \beta_1^{(r)},\pm \beta_3^{(r)},(\cdots),\pm \beta_{2c_r-1}^{(r)},0$, and hence the $r$-th block of the diagonal matrix $D_b=U^{-1}_0(-i)\mathcal{D}bU_0$ is equal to:
\be\label{eqdbs}
D_b^{(r)}(t)={\rm diag}\left(\beta_{1}^{(r)},-\beta_{1}^{(r)},\beta_{3}^{(r)},-\beta_{3}^{(r)},\cdots,\beta_{2c_r-1}^{(r)},-\beta_{2c_r-1}^{(r)},0,\cdots,0\right)\, .
\ee
We define $\beta^{(r)}_\alpha\equiv(D_b^{(r)})_{\alpha\alpha}$ with $1\leq \alpha\leq s_r$. Note that if $\beta^{(r)}_\alpha\neq 0$ there is another $\beta^{(r)}_{\alpha'}=-\beta^{(r)}_\alpha$.
Moreover, with this choice, $(\mathcal{D}b)^2=U_0(iD_b)^2U_0^{-1}=-D_b^2$ is diagonal, and hence $A=U_0(-4D^2)U_0^{-1}=-4D^2$  is also diagonal  with elements  $-4(H^{(r)})^2-(a^{(r)})^{-4}~(\beta^{(r)}_\alpha)^2$.

Since $b(t_0)=\mathbf{0}$, the field $\displaystyle b(t)=\int_{t_0}^t dt'~n(t')~\mathcal{D}b(t')$ takes the form of a block diagonal matrix, with $r$-th block 
\be\label{eqb-antis-blockrs}
b^{(r)}(t)={\rm diag}\left(B_{1}^{(r)}~i\sigma_2,B_{3}^{(r)}~i\sigma_2,\cdots,B_{2c_r-1}^{(r)}~i\sigma_2,\mathbf{0}\right)\, ,
\ee
where
   $\displaystyle B_\alpha^{(r)}(t)\equiv \int_{t_0}^t dt'~n(t')~\beta_\alpha^{(r)}(t')$ are generic eigenvalues of $-ib(t)$. Notice that $b(t)$ commutes with $\mathcal{D}b(t)$, and then $[b,A]=\mathbf{0}$. 

On the other hand, if the solution $U(t)$ to \eqref{equnitaryus} is taken to be  time-dependent,
$\mathcal{D}b$ cannot take the simple form \eqref{eqdb-antis-blockr}, but  $D_b$ can  still  take the form \eqref{eqdbs}. Moreover, in principle  $b(t)$ cannot be as simple as \eqref{eqb-antis-blockrs} and does not commute with $\mathcal{D}b$. However, the information on $U(t)$ is not relevant for the first four equations in \eqref{eqsmotion-6s}, and hence it is not significant for the dynamics of $\Phi$, the scale factors (i.e. the metric) and the eigenvalues of $-i\mathcal{D}b$.

Considering the metric in the $r$-th block  $g^{(r)}=(a^{(r)})^2~\mathbf{1}_{s_r}$, the $\alpha$ component of the equations \eqref{eqa3-6s} and \eqref{eqc1-6s}  may be written as:
\begin{subequations}
\begin{align}
-Q_\alpha^{(r)}/8&=e^{-\Phi}~(a^{(r)})^{-4}~\beta^{(r)}_\alpha~(\mathcal{F}_b'(-4D^2))^{(r)}_{\alpha\alpha}=\text{ constant}\, ,\label{eqa3-8}\\
0&=-\beta^{(r)}_\alpha~Q_\alpha^{(r)}/8+\mathcal{D}\left(e^{-\Phi}~2H^{(r)}(\mathcal{F}_a'(-4D^2))^{(r)}_{\alpha\alpha}\right)\, ,\label{eqc1-8}
\end{align}
\end{subequations}
where  we   defined  a generic eigenvalue of $8~i\mathcal{Q}_1$ as $Q_\alpha^{(r)}\equiv -8(D_{1}^{(r)})_{\alpha\alpha} $ for $1\leq \alpha\leq s_r$. Note that
\eqref{eqc1-8}
 is equivalent to 
\be
-Q'_\alpha\,^{(r)}/8=-B^{(r)}_\alpha~Q_\alpha^{(r)}/8+e^{-\Phi}~2H^{(r)}(\mathcal{F}_a'(-4D^2))^{(r)}_{\alpha\alpha}=\text{ constant}\, ,
\ee
where $-Q'_\alpha\,^{(r)}/8$ and $B_\alpha^{(r)}$ are generic eigenvalues of ${\mathcal{Q}}_2$ and $b$ respectively, if  $U(t)=U_0$.
This is because  ${\mathcal{Q}}_2$ commutes with $\mathcal{D}b$, since $b$ commutes with $\mathcal{D}b$ provided  $U(t)=U_0$, and then it can  be diagonalized in the same basis as $\mathcal{D}b$.
Instead, if  $U(t)$ is a time-dependent  matrix,    $-Q'_\alpha\,^{(r)}/8$ and $B^{(r)}_\alpha$ cannot be interpreted
 as generic eigenvalues of ${\mathcal{Q}}_2$ and $b(t)$, respectively. They  can only be interpreted respectively as an  integration constant related to $\mathcal{Q}_2$ \footnote{Defining $\mathcal{Q}_2^U(t)\equiv U^{-1}(t)\mathcal{Q}_2 U(t)$ and $b^U(t)\equiv U^{-1}(t)b(t) U(t)$,   \eqref{eqc1-4s}  implies
 \be\nonumber
-8(\mathcal{Q}_2^U)_{\alpha\alpha}^{(r)}=Q'_\alpha\,^{(r)}-Q_\alpha^{(r)}~i((b^U)_{\alpha\alpha}^{(r)}-iB_\alpha\,^{(r)})\, , \qquad
-8(\mathcal{Q}_2^U)_{\alpha'\alpha}^{(r)}=-Q_{\alpha}^{(r)}~i(b^U)_{\alpha'\alpha}^{(r)}~~~\text{ for }\alpha'\neq \alpha\, .
  \ee
even for a time dependent $U(t)$. Since $\mathcal{Q}_2,b(t),U(t)$ are  block diagonal matrices, the non-diagonal blocks (with $r\neq r'$)  of $\mathcal{Q}_2, \mathcal{Q}_2^U(t), b(t),b^U(t)$  are zero. 
Note that $Q_\alpha^{(r)}=0\ \forall~(r,\alpha)\implies\mathcal{Q}_1=\mathbf{0}$. If also $Q'_\alpha\,^{(r)}=0\ \forall~(r,\alpha)$, then $\mathcal{Q}_2^U=\mathbf{0}\implies\mathcal{Q}_2=\mathbf{0}$, and finally $\mathcal{Q}=\mathbf{0}$. Conversely, if $\mathcal{Q}=\mathbf{0}$ then $\mathcal{Q}_1=\mathcal{Q}_2=\mathbf{0}$ and $\mathcal{Q}_2^U=\mathbf{0}$, therefore $Q_\alpha^{(r)}=Q'_\alpha\,^{(r)}=0$ for all $(r,\alpha)$.
} and as a primitive of the eigenvalue $\beta^{(r)}_\alpha(t)$ of $-i\mathcal{D}b$. Moreover, we cannot conclude that $b$ and ${\mathcal{Q}}_2$ commute with $\mathcal{D}b$.

Furthermore, defining \be
h_\alpha^{(r)}=(D^{(r)})_{\alpha\alpha}=\pm~\sqrt{(H^{(r)})^2+\frac14(a^{(r)})^{-4}(\beta_\alpha^{(\alpha)})^2}\ee for $1\leq \alpha\leq s_r$ and $1\leq r\leq r_b$, and leaving the sign unspecified for the moment, we get
\be\label{efeh}
\begin{split}
\mathcal{F(DS)}&=\mathcal{F}_a(A)=\mathcal{F}_a(-4D^2)\\
&=8c_1~\displaystyle\sum_{i=1}^d h_i^2+\displaystyle\sum_{k=2}^\infty(-\alpha')\,^{k-1}~2^{2k}~\displaystyle\sum_{P\in~\text{Part}(k,2)}2^{|P|}c_{k,P}\prod_{m\in P}\sum_{i=1}^d h_i^{2m}\\
&\equiv F_h(h_1,h_2,\cdots,h_d)
\end{split}
\ee
as a multi-variable function of the eigenvalues of $A$ ($h_\alpha^{(r)}$ are relabeled $h_i$ for $1\leq i\leq d$ since  $\displaystyle\sum_{r=1}^{r_b}\sum_{\alpha=1}^{s_r}=\displaystyle\sum_{i=1}^d=d$). 

In addition, $\mathcal{F}_a'(-4D^2)$
is a real diagonal matrix with elements 
\be
\begin{split}
&(\mathcal{F}_a'(-4D^2))_{ii}=\\
&=\displaystyle\frac{-1}{8h_i}\left(16c_1h_i+\displaystyle\sum_{k=2}^\infty(-\alpha')^{k-1}2^{2k}\displaystyle\sum_{P\in~\text{Part}(k,2)}2^{|P|}c_{k,P} \displaystyle\sum_{m_0\in P}2m_0h_i^{2m_0-1}\prod_{m\in P-\{m_0\}}\displaystyle\sum_{i'=1}^d h_{i'}^{2m}\right)\\
&=-\frac{1}{8h_i}~\frac{\partial F_h(h_1,h_2,\cdots,h_d)}{\partial h_i}\, .
\end{split}
\ee
Notice that if $h_i=0$, or equivalently if $H_i$ and $\beta_i$ are both zero, this expression is equal to $-2c_1$.
Then
\be
\tr(\mathcal{DS~F}'(\mathcal{DS}))=2\tr(A\mathcal{F}_a'(A))=2\tr((-4D^2)~\mathcal{F}_a'(-4D^2))=\displaystyle\sum_{i=1}^d h_i \frac{\partial F_h}{\partial h_i}\, ,
\ee
and  the diagonalized equations of motion can  be rewritten in terms of the multi-variable function $F_h(h_1,\cdots,h_d)$ and its first order partial derivatives as
\begin{subequations}
\begin{align}
Q_\alpha^{(r)}&=e^{-\Phi}~(a^{(r)})^{-4}~\beta^{(r)}_\alpha~\displaystyle\frac{1}{h^{(r)}_{\alpha}}\frac{\partial F_h}{\partial h^{(r)}_{\alpha}}=\text{ constant}\, ,\label{eqa3-9}\\[0.15cm]
0&=Q_\alpha^{(r)}~\beta^{(r)}_\alpha+\mathcal{D}\left(e^{-\Phi}~2H^{(r)}~\displaystyle\frac{1}{h^{(r)}_{\alpha}}\frac{\partial F_h}{\partial h^{(r)}_{\alpha}}\right)\, ,\label{eqc1-9}\\[0.15cm]
E_n&=(\mathcal{D}\Phi)^2-F_h(h_1,\cdots,h_d)+\displaystyle\sum_{r=1}^{r_b}\displaystyle\sum_{\alpha=1}^{s_r} h^{(r)}_{\alpha} \frac{\partial F_h}{\partial h^{(r)}_{\alpha}}=0\, ,\label{eqEnFh}\\[0.15cm]
E_\Phi+E_n&=2\mathcal{D}^2\Phi+\displaystyle\sum_{r=1}^{r_b}\displaystyle\sum_{\alpha=1}^{s_r} h^{(r)}_{\alpha} \frac{\partial F_h}{\partial h^{(r)}_{\alpha}}=0\, ,\label{eqEphiFh}
\end{align}
\end{subequations}

Note that the second equation is equivalent to
\be \label{eqc1-9int}
Q'_\alpha\,^{(r)}=Q_\alpha^{(r)}~B^{(r)}_\alpha+e^{-\Phi}~2H^{(r)}~\displaystyle\frac{1}{h^{(r)}_{\alpha}}\frac{\partial F_h}{\partial h^{(r)}_{\alpha}}=\text{ constant}\, ,
\ee
in terms of the primitive $B^{(r)}_\alpha$ 
of $\beta^{(r)}_\alpha$.

\subsection{Conservation of the Noether charge and $b$-field dynamics }\label{subsec-cons-Q}

To analyze the conditions for the conservation of the Noether charge $\mathcal{Q}$,  recall that 
they are equivalent to the conservation of all the scalar Noether charges $Q_{\alpha}^{(r)}$ and $Q'^{(r)}_{\alpha}$, together with the condition \eqref{equnitaryus} for $U(t)$ (which we may ignore from now on, since it does not influence the other equations).
We will consider only one diagonal component $i$ (or $r,\alpha$) and  deal separately with  the cases $i$) $Q_{\alpha}^{(r)}=Q'^{(r)}_{\alpha}= 0$; $ii$) $Q_{\alpha}^{(r)}=0$, $Q'^{(r)}_{\alpha}\neq 0$ and $iii$) $Q_{\alpha}^{(r)}\neq 0$.

  \noindent $i$) $Q_{\alpha}^{(r)}=0$ and $Q'_{\alpha}\,^{(r)}= 0$:
    
    In this case, the equations \eqref{eqa3-9} and \eqref{eqc1-9int}  imply  
\be  0=\pm\left[(H^{(r)})^2+\frac14(a^{(r)})^{-4}(\beta^{(r)}_\alpha)^2\right]~\displaystyle\frac{1}{h^{(r)}_{\alpha}}\frac{\partial F_h}{\partial h^{(r)}_{\alpha}}=h^{(r)}_{\alpha}\frac{\partial F_h}{\partial h^{(r)}_{\alpha}} \, ,\label{eqhrdfh4}
\ee and then either $\displaystyle\frac{\partial F_h}{\partial h^{(r)}_{\alpha}}=0$ or  $h^{(r)}_{\alpha}=0 \iff H^{(r)}=\beta_\alpha^{(r)}=0$. Moreover, the solution $h^{(r)}_{\alpha}=0$ is contained in the equation \be\label{eqder0-case-i}\displaystyle\frac{\partial F_h}{\partial h^{(r)}_{\alpha}}=0 \, , \ee and then this equation is equivalent to \eqref{eqa3-9} and \eqref{eqc1-9int}.  We can choose  $\pm=\sign(h^{(r)}_{\alpha})=\sign(H^{(r)})$, or if $H^{(r)}=0$ we can choose $\pm=\sign(h^{(r)}_{\alpha})=\sign(\beta^{(r)}_{\alpha})$, without loss of generality.

 \noindent $ii$) $Q_{\alpha}^{(r)}=0$ and $Q'_{\alpha}\,^{(r)}\neq 0$: 
\be
Q'_{\alpha}\,^{(r)}\neq 0\implies \displaystyle\frac{1}{h^{(r)}_{\alpha}}\frac{\partial F_h}{\partial h^{(r)}_{\alpha}}\neq 0\ \ {\rm and \ then}\ \ Q_{\alpha}^{(r)}=0\implies \beta_{\alpha}^{(r)}=0\, ,
\ee
which  trivially verifies \eqref{eqa3-9}.     
    On the other hand,  since $\beta_{\alpha}^{(r)}=0\implies h^{(r)}_{\alpha}=H^{(r)}$ (choosing $\pm=\sign(h^{(r)}_{\alpha})=\sign(H^{(r)})$), \eqref{eqc1-9int} implies $h^{(r)}_{\alpha}=H^{(r)}\neq 0$ and takes the form
    \begin{equation}
   \frac12Q'_\alpha\,^{(r)}\equiv q^{(r)}_{\alpha}= e^{-\Phi}~\frac{\partial F_h}{\partial h^{(r)}_{\alpha}}=\text{ constant}\neq 0\, .
    \end{equation}

  \noindent $iii$) $Q_{\alpha}^{(r)}\neq 0$: 

In this case $\displaystyle\frac{1}{h^{(r)}_{\alpha}}\frac{\partial F_h}{\partial h^{(r)}_{\alpha}}\neq 0$ and  $\beta^{(r)}_\alpha\neq 0$ for all times.
    Hence,  \eqref{eqa3-9} can be rewritten as:
    \be
    e^{-\Phi}~\displaystyle\frac{1}{h^{(r)}_{\alpha}}\frac{\partial F_h}{\partial h^{(r)}_{\alpha}}=Q_{\alpha}^{(r)}~\displaystyle\frac{(a^{(r)})^4}{\beta^{(r)}_\alpha}
    \ee
 Replacing this in \eqref{eqc1-9int}, this equation is equivalent to:
    \begin{equation}
    \begin{split}
    &B_{\alpha}^{(r)} +\displaystyle\frac{1}{2}\frac{\mathcal{D}((a^{(r)})^4)}{\mathcal{D}(B_{\alpha}^{(r)})}=\displaystyle\frac{Q'_\alpha\,^{(r)}}{Q_{\alpha}^{(r)}}\equiv q'_{\alpha}\,^{(r)} =\text{ constant}\\
    \iff&\mathcal{D}\left((a^{(r)})^4+(B_{\alpha}^{(r)})^2-2q'_{\alpha}\,^{(r)}B_{\alpha}^{(r)} \right)=\mathcal{D}\left((a^{(r)})^4+\left(B_{\alpha}^{(r)}-q'_{\alpha}\,^{(r)}\right)^2 \right)=0\, ,
    \end{split}
    \end{equation}
   which can be expressed in terms of a non-negative integration constant $R_r^2$ as:
    \begin{equation}
    (a^{(r)})^4+\left(B_{\alpha}^{(r)}-q'_{\alpha}\,^{(r)}\right)^2 =R_r^2=\text{constant} \geq 0\, ,
    \end{equation}
    or equivalently as:
    \begin{equation}\label{eqabetacostheta}
    \begin{cases}(a^{(r)})^2=R_r \cos(\theta^{(r)}_\alpha(t)) \implies 2H^{(r)}(a^{(r)})^2=-R_r \sin(\theta^{(r)}_\alpha(t))~\mathcal{D}\theta^{(r)}_\alpha(t)\\
    B_{\alpha}^{(r)}-q'_{\alpha}\,^{(r)}=R_r \sin(\theta^{(r)}_\alpha(t))\implies \beta_{\alpha}^{(r)}=R_r \cos(\theta^{(r)}_\alpha(t))~\mathcal{D}\theta^{(r)}_\alpha(t)\\
    \end{cases}
    \end{equation}
    with $\theta^{(r)}_\alpha(t)\in (-\pi/2,\pi/2)$ so that $(a^{(r)})^2\geq 0$. The fact that $\beta_{\alpha}^{(r)}\neq 0$ implies that $R_r\neq 0$,  $a^{(r)}(t)\neq 0$ and  $\mathcal{D}\theta^{(r)}_\alpha(t)\neq 0$. 
    
    Notice that $(a^{(r)}(t))^2\leq R_r <\infty$ is bounded, and then this  cannot correspond to a dS cosmology in the string frame in the $s_r$ spatial directions of the $r$-th block.
     This is a rather curious feature of the dynamics of the system when the $b$-field is turned on in this case: the scale factor $a^{(r)}$ corresponding to  a non-trivial eigenvalue $\beta_{\alpha}^{(r)}(t)$ of $\mathcal{D}b$  is bounded.
    
    Furthermore, using \eqref{eqabetacostheta} and choosing $\pm=\sign(h^{(r)}_\alpha)=-\sign(\mathcal{D}\theta^{(r)}_\alpha)$, we see that:
    \be \nonumber
    H^{(r)}=-\displaystyle\frac{1}{2}\tan(\theta_{\alpha}^{(r)}(t))\mathcal{D}(\theta_{\alpha}^{(r)}(t)) \ \qquad {\rm and} \  \qquad h^{(r)}_\alpha=-\displaystyle\frac{\mathcal{D}\theta_{\alpha}^{(r)}}{2~\cos(\theta_{\alpha}^{(r)})}\, ,
    \ee
    hence $\pm=\sign(h^{(r)}_\alpha)=-\sign(\beta^{(r)}_\alpha)$ and
    \be
    \displaystyle\frac{h_{\alpha}^{(r)}~(a^{(r)})^4}{\beta^{(r)}_\alpha}=h_{\alpha}^{(r)}~(a^{(r)})^2~\displaystyle\frac{(a^{(r)})^2}{\beta^{(r)}_\alpha}=-\displaystyle\frac{\mathcal{D}\theta_{\alpha}^{(r)}}{2}~R_r~\displaystyle\frac{1}{\mathcal{D}\theta_{\alpha}^{(r)}} =-\displaystyle\frac{R_r}{2}\,.
    \ee
  Thus, turning back to  \eqref{eqa3-9}, we get:
    \begin{equation}
    e^{-\Phi}~\frac{\partial F_h}{\partial h^{(r)}_{\alpha}}=-\displaystyle\frac{R_r~Q_\alpha^{(r)}}{2}\equiv q^{(r)}_{\alpha}=\text{ constant}\neq 0\, .
    \end{equation}

Summarizing, in the generalized ansatz, the (diagonalized) equations of motion are:
\begin{subequations}
\begin{align}
&E_n=(\mathcal{D}\Phi)^2-F_h(h_1,\cdots,h_d)+\displaystyle\sum_{r=1}^{r_b}\displaystyle\sum_{\alpha=1}^{s_r} h^{(r)}_{\alpha} \frac{\partial F_h}{\partial h^{(r)}_{\alpha}}=0\, ,\label{eqEnFh-4.2}\\[0.15cm]
&E_\Phi+E_n=2\mathcal{D}^2\Phi+\displaystyle\sum_{r=1}^{r_b}\displaystyle\sum_{\alpha=1}^{s_r} h^{(r)}_{\alpha} \frac{\partial F_h}{\partial h^{(r)}_{\alpha}}=0\, ,\label{eqEphiFh-4.2}\\[0.15cm]
&e^{-\Phi}\frac{\partial F_h}{\partial h^{(r)}_{\alpha}}= q^{(r)}_{\alpha}=\text{ constant}\label{eqcargamulti}\\
&q^{(r)}_{\alpha}\neq 0\implies\begin{cases} \text{case $ii$): }Q_{\alpha}^{(r)}=0 \text{ and } Q'_{\alpha}\,^{(r)}\neq 0~\implies~\beta^{(r)}_{\alpha}=0 \\
\text{case $iii$): }Q_{\alpha}^{(r)}\neq 0\implies\begin{cases}h^{(r)}_\alpha=-\displaystyle\frac{\mathcal{D}\theta_{\alpha}^{(r)}}{2~\cos(\theta_{\alpha}^{(r)})}\\(a^{(r)})^2=R_r \cos(\theta^{(r)}_\alpha(t)) 
\\
B_{\alpha}^{(r)}-q'_{\alpha}\,^{(r)}=R_r \sin(\theta^{(r)}_\alpha(t))
\end{cases}
\end{cases}\label{eqqneq0}
\end{align}
\end{subequations}
where \eqref{eqcargamulti} and \eqref{eqqneq0} for every pair of indexes $(r,\alpha)$ are equivalent to the (diagonalized) equation of motion for $\mathcal{S}$ variations.

The equation \eqref{eqcargamulti} for all $(r,\alpha)$, together with \eqref{eqEnFh-4.2} and \eqref{eqEphiFh-4.2}, determine the dynamics of the parameters $h^{(r)}_{\alpha}(t)$ and the generalized dilaton $\Phi$.

In addition, the equation \eqref{eqqneq0} establishes $\beta^{(r)}_{\alpha}=0 $ and $H^{(r)}=h^{(r)}_{\alpha}$  in the case $ii$); while in the case $iii$), it determines $\theta^{(r)}_{\alpha}(t)$ with a first order differential equation in terms of $h^{(r)}_{\alpha}(t)$, which in turn fixes $(a^{(r)})^2,~B_{\alpha}^{(r)}-q'_{\alpha}\,^{(r)}$ and hence $H^{(r)}(t)$ and $\beta^{(r)}_{\alpha}(t)$.

In the case $q^{(r)}_{\alpha}=0$, or equivalently  $Q^{(r)}_{\alpha}=Q'\,^{(r)}_{\alpha}=0$, i.e. case $i$), the component $(r,\alpha)$ of the diagonalized equation of motion for $\mathcal{S}$ is equivalent only to the equation $\displaystyle\frac{\partial F_h}{\partial h^{(r)}_{\alpha}}= 0$.

If  the equations of motion are considered  perturbatively up to $\mathcal{O}(\alpha'\,^0)$, then $F_h(\vec{h})=8c_1\displaystyle\sum_{i=1}^dh_i^2+\mathcal{O}(\alpha')$, and hence \eqref{eqcargamulti} implies $(e^{-\Phi}h_\alpha^{(r)})_0=\displaystyle\frac{(q_{\alpha}^{(r)})_0}{16c_1}=$constant. To construct a perturbative dS solution in at least one component (i.e. $H^{(r)}=H_0^{(r)}\neq 0$ for at least one $r$), one needs $Q_{\alpha}^{(r)}= 0$ for every $1\leq \alpha\leq s_r$ (to avoid a bounded scale factor), and  $(Q'_{\alpha}\,^{(r)})_0=2(q_{\alpha}^{(r)})_0\neq 0$ (to avoid a Minkowski solution with $(H_0^{(r)})_0=0$). We are left with case $ii)$, in which $\beta_{\alpha}^{(r)}=0,~ h_{\alpha}^{(r)}=H^{(r)}\neq 0$. The equation $e^{-\Phi_0(t)}(H^{(r)})_0=(q^{(r)}_\alpha)_0=$constant only allows a non-zero constant $(H^{(r)})_0$ if $\Phi_0(t)$ is constant, but equation \eqref{eqEphiFh} implies  $0=|2(\mathcal{D}^2\Phi)_0|=\left|\displaystyle\sum_{r=1}^{r_b}\sum_{\alpha=1}^{s_r} \left(h^{(r)}_{\alpha} \frac{\partial F_h}{\partial h^{(r)}_{\alpha}}\right)_0\right|=16|c_1|~\displaystyle\sum_{r=1}^{r_b}\sum_{\alpha=1}^{s_r} (h^{(r)}_{\alpha})_0^2 \implies (h^{(r)}_{\alpha})_0=0\implies (H^{(r)})_0=0$, then there is no solution of the two-derivative equations with non-zero constant Hubble parameter. Therefore, we see that  in the generalized ansatz, the theory does not allow perturbative dS solutions up to $\mathcal{O}(\alpha'\,^0)$, not even in one spatial component. Although we only proved this up to leading order, it makes sense that there are no perturbative dS solutions to all orders since $H_i=$constant should have units of $1/\sqrt{\alpha'}$, and hence it would not be perturbative.  This is why we  now turn to search for non-perturbative solutions.

In a similar fashion as in the isotropic ansatz considered in the preceding section, the sector with vanishing Noether charge contains many interesting non-perturbative dS solutions, both with vanishing or non-vanishing $b$-field. Indeed, we show in  appendix \ref{app:qne0} that if every $h_i$ is constant, which is a key for the construction of dS solutions, then $\mathcal{Q}=\mathbf{0}$.

\section{Generalized non-perturbative dS vacua} \label{sec:bneq0}

To search for dS solutions in the case $\mathcal{Q}=\mathbf{0}$, we recall the equations of motion \eqref{eomf}, copied here for convenience
\begin{subequations}\label{eomf3}
\begin{align}
&E_n=(\mathcal{D}\Phi)^2-\mathcal{F(DS)}+\text{tr}\left[\mathcal{DS}~\mathcal{F}'(\mathcal{DS})\right]=0\, ,\label{eomfn3}\\
&E_\Phi+E_n=2\mathcal{D}^2\Phi+\text{tr}\left[\mathcal{DS}~\mathcal{F}'(\mathcal{DS})\right]=0\, ,\label{eomfphi3}\\
& E_\mathcal{S}=-\frac{1}{2}e^\Phi\mathcal{S}~\mathcal{DQ}=\mathbf{0} \iff \mathcal{DQ}=\mathbf{0}\iff \mathcal{Q}=\text{constant}\in\mathfrak{so}(d,d)\, .
\end{align}
\end{subequations}
The condition $\mathcal{Q}=-2~e^{-\Phi}~\mathcal{S}~\mathcal{F'(DS)}=\mathbf{0}$ necessarily implies that $\mathcal{F}'(\mathcal{DS})=\mathbf{0}$  for all times when $e^{-\Phi}\neq 0$ and  $\mathcal{S}$ is invertible. Then  \eqref{eomfphi3}  implies  $\mathcal{D}^2\Phi=0 \iff \mathcal{D}\Phi=-c=$ constant $\in \mathbb{R}$ and \eqref{eomfn3} requires $\mathcal{F(DS)}=c^2$. 
Summarizing, the generic solutions with null Noether charge must satisfy:
\be\label{eqgenericds}
\mathcal{F}'(\mathcal{DS})=\mathbf{0}~~,~~\mathcal{F(DS)}=c^2\geq 0~~,~~\mathcal{D}\Phi=-c~~,
\ee
a natural generalization of the conditions \eqref{eqsimpleds}.

Considering that $\mathcal{F(DS)=G(DS}^2)$ only depends on $(\mathcal{DS})^2$ and $\mathcal{F'(DS)}=2~\mathcal{DS}~\mathcal{G}'(\mathcal{DS}^2)$, in principle from \eqref{eqgenericds}
 we can only conclude that $\mathcal{G}'(\mathcal{DS}^2)$ belongs to the matrix subspace that is annihilated when multiplied by $\mathcal{DS}$. 
The  solutions verifying  \eqref{eqgenericds} can be constructed  imposing $(\mathcal{DS})^2=$ constant such that $\mathcal{DS}~\mathcal{G}'(\mathcal{DS}^2)=\mathbf{0}$ and $\mathcal{G}(\mathcal{DS}^2)=c^2\geq 0$.

Imposing $\mathcal{DS}^2=$constant without an ansatz that simplifies the expression \eqref{eqdssqgeneral} seems quite not trivial, since it is necessary to ensure that both the blocks of the first column $A$, $C$ and the blocks of the second column (written in terms of $A$, $C$) are all constant. However, in the generalized ansatz of commuting matrices, this expression takes the simpler form \eqref{eqdssqgenansatz}, and $\mathcal{DS}^2=$constant is equivalent to $A=$constant, which implies that $h_i=$constant.

As we  explained in the previous section,    under the  ansatz of commuting matrices 
the condition $\mathcal{Q}=\mathbf{0}$ is equivalent to ${\mathcal{Q}}_1={\mathcal{Q}}_2=\mathbf{0}$ or  to zero scalar Noether charges $Q_{\alpha}^{(r)}=Q'_{\alpha}\,^{(r)}=0$ for all $(r,\alpha)$. Therefore, as we showed in the case $i)$ of section \ref{subsec-cons-Q}, the equations \eqref{eqa3-9} and \eqref{eqc1-9int} in this case  are equivalent to
$
\displaystyle\frac{\partial F_h}{\partial h_{\alpha}^{(r)}}=0
$
for each $(r,\alpha)$.

The remaining equations of motion $E_n=E_\Phi+E_n=0$  imply  $\mathcal{D}\Phi=-c=$constant $\in\mathbb{R}$ and $F_h(h_1,\cdots,h_d)=c^2$ (see \eqref{eqEnFh} and \eqref{eqEphiFh}). Therefore, the condition \eqref{eqgenericds} for a solution with $\mathcal{Q}=\mathbf{0}$ turns out to be equivalent to
\be\label{eqgenansatzds}
\frac{\partial F_h}{\partial h^{(r)}_{\alpha}}=0 \iff \nabla F_h=\Vec{0}~~,~~F_h(h_1,\cdots,h_d)=c^2\geq 0~~,~~\mathcal{D}\Phi=-c~~,
\ee
for which we require that the parameters $h^{(r)}_{\alpha}$  are constant in order to ensure $\nabla F_h=\Vec{0}$. Again, these solutions seem to be the natural generalization of the conditions \eqref{eqsimpleds}, now written in terms of the multi-variable function $F_h(h_1,\cdots,h_d)$.
These solutions have $h^{(r)}_{\alpha}=$constant, which is not equivalent to $H^{(r)}=$ constant, except for example if $\mathcal{D}b=\mathbf{0}$ where $h^{(r)}_{\alpha}=H^{(r)}$. 

Hence, imposing $H^{(r_0)}=\text{constant}$ for a certain block $r_0$, i.e. a dS solution in the string frame for  such block, requires that $\beta_{\alpha}^{(r_0)}(t) =\text{constant}_\alpha \cdot (a^{(r_0)}(t))^2$, or equivalently $\mathcal{D}b^{(r_0)}(t)=(a^{(r_0)}(t))^2 Y^{(r_0)}$ for the case $U(t)=U_0$, with $Y^{(r_0)}$ a real antisymmetric constant matrix of size $s_{r_0}\times s_{r_0}$. However, for this $\mathcal{Q}=\mathbf{0}$ solution, the condition $(h_\alpha^{(r)})^2=$ constant must be verified for every value of $r$, not only for $r_0$. In principle, some  Hubble parameters might be non-constant  provided  a non-trivial $b$-field compensates the time dependence of $H^{(r)}(t)$ and the dimension $s_r$ of the block is even, since  $D_b^{(r)}$ cannot have a zero element. Of course, if we are not interested in this case, we can simply take $H^{(r)}=$ constant for every $r$.

\subsection{dS solutions in the Einstein frame and examples}

As discussed in section \ref{subsec:eins}, dS solutions in the Einstein frame can be obtained from the dS solutions in the string frame if the dilaton $\phi=\frac12[\Phi+\ln(\sqrt{\det{g_{ij}}})]$ is constant.
For a diagonal metric $g_{ij}=a_i^2(t)~\delta_{ij}$, \eqref{eqgenericds} requires
\be\label{eqdilconstant}
\mathcal{D}(\ln(\sqrt{\det{g_{ij}}}))=\displaystyle\sum_{i=1}^d H_i =c=\text{ constant }\in\mathbb{R} \, .
\ee
In a solution with  non-zero constant Hubble parameters  only in 
 $n<d$ spatial dimensions, i.e. $H_i=H_{0,i}=\text{ constant }\neq 0$ for $i=1,\cdots,n$, this becomes $\displaystyle\sum_{i=n+1}^d H_i=c-\displaystyle\sum_{i=1}^nH_{0,i}=$constant. 
Note that  the remaining Hubble parameters which are not a non-zero constant may have a temporal dependence provided they add up to a constant, i.e. the temporal dependence must cancel in the sum. This cannot occur if the metric is isotropic in the extra $d-n$ spatial dimensions, in which case the geometry of those extra dimensions corresponds to a static cosmology, and the condition \eqref{eqgenericds} for a constant dilaton takes the form $c=\displaystyle\sum_{i=1}^nH_{0,i}$.

If the constants $h_i=h^{(r)}_\alpha=h_{0,i}$ that solve the equation \eqref{eqgenansatzds} for a certain value of $c$ are known,  some interesting particular dS solutions can be constructed, as we now discuss.

\subsubsection{Isotropic dS geometry and $b\ne\mathbf{0}$} 

 Consider an isotropic dS geometry in $n$ spatial dimensions and a static solution in the remaining $d-n$ spatial dimensions, i.e. the metric is $g=\diag_n(a^2(t),a_0^2)$, with $a(t)$ such that $H=\mathcal{D}(\ln{a(t)})=H_0=$ constant and $a_0=$ constant.  Then, there are two blocks: 
    \begin{enumerate}
        \item The $r=1$ block of size $n\times n$ with $g^{(1)}=a^2(t)~\mathbf{1}_n$, $H^{(1)}=H_0=$constant $\neq 0$ and $\mathcal{D}b^{(1)}=a^2(t)Y^{(1)}$ with $Y^{(1)}=$constant.
        \item The $r=2$ block of size $(d-n)\times (d-n)$ with $g^{(2)}=a^2_0~\mathbf{1}_{d-n}=$constant, $H^{(2)}=H_2=0$ and $\mathcal{D}b^{(2)}=a^2_0Y^{(2)}=$constant.
    \end{enumerate} 
    Assuming $U(t)=U_0$ for simplicity, the temporal dependence of $\mathcal{D}b^{(r)}(t)=(a^{(r)}(t))^2Y^{(r)}$ is necessary to have $h^{(r)}_\alpha=$constant. These constants must verify \eqref{eqgenansatzds} to be a $\mathcal{Q}=\mathbf{0}$ solution.

   If  the condition for a constant dilaton  $c=nH_0$  is fulfilled, the solution is $dS_n\times T^{d-n}$  in both frames. The difference with the previous solution of \eqref{eqsimpleds} is that now it allows a non-trivial $b$-field.  In particular, if $n=d$ this is a dS solution, isotropic in all the $d$ spatial dimensions with $H=H_0=$ constant, and a non-trivial $b$-field such that $\mathcal{D}b=a^2(t)Y^{(1)}$ for which the condition that $\mathcal{D}b$ is block diagonal is always verified, since there is only one block.

\subsubsection{Anisotropic dS geometry and $b=\mathbf{0}$}

If  $b=\mathbf{0}$, then  $h_i=H_i$.     
    A solution with $\mathcal{Q}=\mathbf{0}$ is  simply obtained when the $h_i=h_{0,i}$ are the constants  that solve the  equation \eqref{eqgenansatzds}. This corresponds to an anisotropic dS solution in all the $n$ spatial dimensions  and a static geometry in the remaining $d-n$ spatial dimensions.  Moreover, if $c=\displaystyle\sum_{i=1}^nH_{i,0}$, this anisotropic dS geometry has a constant dilaton, and it is dS in both frames.
    
\subsection{Stability of dS solutions}

Generalizing the analysis of stability performed in \ref{subsec:stability}, we define $y\equiv\mathcal{D}\Phi$ and recall that the equations of motion  \eqref{eqcargamulti} and \eqref{eqsmotion-6snphi} are first order differential equations for $y$ and for all the parameters $h_i$ respectively, while  \eqref{eqsmotion-6sn} is a constraint between them.
Therefore, a variation of the dynamical variables $\delta y=\delta(\mathcal{D}\Phi)$ and $\delta h_i$  must preserve the constraint:
\be
\begin{split}
0&=2y~\delta y-\displaystyle\sum_{i=1}^{d}\frac{\partial F_h}{\partial h_i}~\delta h_i+\displaystyle\sum_{i=1}^{d} \left(\delta h_i \frac{\partial F_h}{\partial h_i}+h_i \displaystyle\sum_{i'=1}^{d}\frac{\partial^2 F_h}{\partial h_{i'}\partial h_i}~\delta h_{i'}\right)\\
&=-2c~\delta y+\displaystyle\sum_{i=1}^{d} \displaystyle\sum_{i'=1}^{d}h_{i}~\frac{\partial^2 F_h}{\partial h_{i'}\partial h_i}~\delta h_{i'}\, ,\\
\end{split}
\ee
where everything  is evaluated in the $\mathcal{Q}=\mathbf{0}$ solution, except the variations; for example: $y=-c$ and $h_{i}=h_{i,0}$.

Performing a variation in $E_\Phi+E_n=0$, we get
\be
\mathcal{D}(\delta y)=-\frac{1}{2}\displaystyle\sum_{i=1}^{d}  \displaystyle\sum_{i'=1}^{d}h_i~\frac{\partial^2 F_h}{\partial h_{i'}\partial h_i}~\delta h_{i'}=-c~\delta y\, ,
\ee
 evaluating in the $\mathcal{Q}=\mathbf{0}$ solution and taking $\displaystyle\frac{\partial F_h}{\partial h^{(r)}_{\alpha}}=0$. 
Hence, the dynamics of $y=\mathcal{D}\Phi$ is stable in the $\mathcal{Q}=\mathbf{0}$ solution if $c>0$, and is unstable if $c<0$.

Turning now to the equations \eqref{eqcargamulti} that determine the dynamics of $h_i$, namely:
\be
0=\mathcal{D}\left(e^{-\Phi}\displaystyle\frac{\partial F_h}{\partial h_i}\right) \iff 0=-(\mathcal{D}\Phi)\displaystyle\frac{\partial F_h}{\partial h_i}+\mathcal{D}\left(\displaystyle\frac{\partial F_h}{\partial h_i}\right)=-(\mathcal{D}\Phi)\displaystyle\frac{\partial F_h}{\partial h_i}+\displaystyle\sum_{i'=1}^d\displaystyle\frac{\partial^2 F_h}{\partial h_{i'}\partial h_i}\mathcal{D}h_{i'}
\ee
and performing the variations $\delta y$ and $\delta h_i$, we have:
\be
0=-\delta y~\displaystyle\frac{\partial F_h}{\partial h_i}-y~ \delta\left(\displaystyle\frac{\partial F_h}{\partial h_i}\right)+\mathcal{D}\left(\delta\left(\displaystyle\frac{\partial F_h}{\partial h_i}\right)\right)
\iff  \mathcal{D}\left(\delta\left(\displaystyle\frac{\partial F_h}{\partial h_i}\right)\right)=-c~\delta\left(\displaystyle\frac{\partial F_h}{\partial h_i}\right)\label{prev}
\ee
where  $\displaystyle\frac{\partial F_h}{\partial h_i}=0$ was evaluated in the $\mathcal{Q}=\mathbf{0}$ solution.

Hence, the dynamics of every partial derivative $\displaystyle\frac{\partial F_h}{\partial h_i}$ is stable in the $\mathcal{Q}=\mathbf{0}$ solution if $c>0$, and is unstable if $c<0$.

Moreover,  \eqref{prev} can be written as:
\be
\begin{split}
&\displaystyle\sum_{i'=1}^d\left(\displaystyle\frac{\partial^2 F_h}{\partial h_{i'}\partial h_i}\mathcal{D}(\delta h_{i'})+\mathcal{D}\left(\displaystyle\frac{\partial^2 F_h}{\partial h_{i'}\partial h_i}\right)\delta h_{i'}\right)=-c~\displaystyle\sum_{i'=1}^d\displaystyle\frac{\partial^2 F_h}{\partial h_{i'}\partial h_i}\delta h_{i'}\\
&\iff \displaystyle\sum_{i'=1}^d\displaystyle\frac{\partial^2 F_h}{\partial h_{i'}\partial h_i} \Big(\mathcal{D}(\delta h_{i'})+c~\delta h_{i'}\Big)=0\iff \mathcal{D}(\delta h_{i'})=-c~\delta h_{i'}\, ,
\end{split}
\ee
where in the second line we used that $\mathcal{D}\left(\displaystyle\frac{\partial^2 F_h}{\partial h_{i'}\partial h_i}\right)=0$, since  $h_i=$constant, and  assumed that the Hessian matrix of $F_h$ is invertible.
Then, the dynamics of each parameter $h_i$ is stable in the $\mathcal{Q}=\mathbf{0}$ solution if $c>0$, and is unstable if $c<0$.

Therefore, a (possibly dS) $\mathcal{Q}=\mathbf{0}$ solution described by $(h_{0,1},\cdots,h_{0,d},c)$ is stable if $c>0$ and unstable if $c<0$, for  the dynamics of both $\Phi$ and $h_i$. In the case $c=0$, further analysis is required. In particular, a $\mathcal{Q}=\mathbf{0}$ solution with constant dilaton is stable if $c=\displaystyle\sum_{i=1}^d H_i=\text{constant}>0$.

Notice that the time-reversal symmetry always allows to obtain a stable dS solution from a solution with $c\neq 0$, since it transforms $(h_{0,1},\cdots,h_{0,d},c)$ to $(-h_{0,1},\cdots,-h_{0,d},-c)$. Moreover, if $\vec{h}_0=(h_{0,1},\cdots,h_{0,d})$ solves \eqref{eqgenansatzds}, then any other of the possible $2^d-1$ vectors $(\pm h_{0,1},\cdots,\pm h_{0,d})$ obtained from changing some signs of the components also solves it. Hence, one can always choose a dS solution $(h_{0,1},\cdots,h_{0,d},c)$ that is stable and expanding in some  directions and contracting in the others, provided these directions are not static.

\section{Caveats on non-perturbative dS solutions}\label{sec:caveats}

In this section we summarize the procedure to obtain non-perturbative dS solutions, and in particular those with constant dilaton. We also discuss the obstructions to determine whether there are dS solutions or not,  if only  the asymptotic expansion of the function $F_n(H)$ is available. For the sake of clarity, we present the arguments in the simpler case of an isotropic dS geometry with $b=0$ analyzed in section \ref{sec:feq}, and the extension to the generalized case is presented in appendix \ref{app:caveats}.

If  the function $F_n(H)$ is   defined for non-infinitesimal values of $\sqrt{\alpha'}H$ and all the coefficients $c_k^{(n)}$ are known, then it contains non-perturbative information of the theory. As discussed in the previous sections, in this case the theory admits non-perturbative dS solutions if $F'_n(H)=0$, $F_n(H)=c^2\ge 0$ and ${\mathcal D}\Phi=-c$.
To explicitly find these solutions, and especially to determine if they admit  a constant dilaton, one should implement the following steps:
\begin{enumerate}
    \item Calculate the roots $H_0\neq 0$ of $F_n'$. Since $F_n(H_0)=F_n(-H_0)\implies F'_n(H_0)=-F_n'(-H_0)$, given a root $H_0$ there will be another one $-H_0$, and then one can choose only the positive roots $H_0>0$,  corresponding to expanding cosmologies in the string frame.
    \item Keep only the roots $H_0$ such that $F_n(H_0)=c^2\geq 0$ gives a non-negative number.
    \item For each of these $H_0$ values there is a non-perturbative dS solution in the string frame like \eqref{eqsimpleds} if  $\mathcal{D}\Phi=\mp \sqrt{F_n(H_0)}=\mp |c|=-c$, for any choice of $\sign(c)$. Then, assuming $c\neq 0$, there are two dS solutions: a stable one for $c>0$ and an unstable  one  for $c<0$. The stable solution  is an expanding  dS metric in the string frame, corresponding to the green quadrant $(H_0>0,c>0)$ in Figure \ref{ds_str}. Then it is also an expanding  solution in the Einstein  frame with $H_E(t)>0$ if $n<d$ or if $c>H_0$ (cyan and light green regions in Figure \ref{ds_str_eins}). 
    \item In particular, if $c=nH_0 \iff \mathcal{D}\phi=0$, there is a non-perturbative dS solution with constant dilaton $\phi$, which  can be taken to be stable and expanding in both frames. 
\end{enumerate}

To illustrate the procedure, take as an example (not connected with string theory): $F_n(H)=F_0\cos(\sqrt{\alpha'}H)$ with $F_0=\tilde{F}_0/\alpha'>0$ a dimensionful constant. The previous steps become:
\begin{enumerate}
    \item $F_n'(H_0)=-F_0\sqrt{\alpha'} \sin(\sqrt{\alpha'}~H_0)=0\implies
    \sqrt{\alpha'}~H_0=m_1\pi\iff H_0=H_{0,m_1}=\displaystyle\frac{m_1\pi}{\sqrt{\alpha'}}
   $
    with $m_1>0$ positive integer since we only keep the solutions with $H_0>0$.
    \item Keep the roots $H_0=H_{0,m_1}$ such that
    \be \nonumber F_n(H_0)=F_0\cos(m_1\pi)=F_0~(-1)^{m_1}=c^2\geq 0 \iff(-1)^{m_1}=+1\, ,\ee i.e.  keep  only the roots with $m_1=2m$ even and positive. 
    \item For each of these $H_{0}=H_{0,2m}$ values, there are two dS solutions with $\mathcal{D}\Phi=-c$: a stable one with $c=+\sqrt{F_0}$ and an unstable one with $c=-\sqrt{F_0}$. Choosing the former, for each $m\in \mathbb{N}$ there is a stable and expanding  solution in the string frame  described by $\left(H_{0,2m}=\displaystyle\frac{2m\pi}{\sqrt{\alpha'}},c=+\sqrt{F_0}\right)$. If $n<d$ or if $c>H_0$, it is also expanding in the Einstein frame.

    \item In particular, if $F_0=n^2H_{0,2m_\phi}^2$ for a certain $m_\phi\in \mathbb{N}$, there is a non-perturbative dS solution $\left(H_{0,2m_\phi}=\displaystyle\frac{2m_\phi\pi}{\sqrt{\alpha'}},c=+\sqrt{F_0}=nH_{0,2m_\phi}\right)$ with constant dilaton $\phi$, which can be taken to be stable and expanding in both frames.
\end{enumerate}

Instead, if  the only available information is the asymptotic expansion of $F_n(H)$, it is not possible to determine whether there are non-perturbative dS solutions or not.  To see this, suppose  that only the values of all the coefficients $c^{(n)}_{k}$ are known in the perturbative expansion and $\tilde{F}_n(\tilde{x})\equiv \alpha' F_n(H)$ is a dimensionless function of the dimensionless variable $\tilde{x}\equiv \sqrt{\alpha'} H $.  
In this case, one cannot distinguish between $\tilde{F}_n(\tilde{x})$
and other functions with the same asymptotic expansion around $\tilde{x}\sim 0$, say $\tilde{F}_n(\tilde{x})+\tilde{h}(\tilde{x})$  with $\tilde{h}(\tilde{x}\sim0)\sim 0$.  In other words,  non-perturbative information of the theory is necessary to distinguish between perturbatively equivalent functions that belong to the same equivalence class \be[\tilde{F}_n]=\{\tilde{F}_n(\tilde{x})+\tilde{h}(\tilde{x}):\tilde{h}(\tilde{x}\sim 0)\sim 0\}\, .\ee
A function $\tilde{h}(\tilde{x})$ with trivial asymptotic expansion $\tilde{h}(\tilde{x})\sim 0$ is said to be subdominant \cite{murrayasympt}:  it decays faster than any polynomial $\tilde{x}^n$ when   $\tilde{x}\sim 0$. Since $\tilde{F}_n(\tilde{x})$ is even (i.e. it only depends on $\tilde{x}^2\propto \alpha'\tr(\mathcal{DS}^2)$, hence it preserves the duality and time-reversal symmetries), the subdominant functions must also be even: $\tilde{h}(\tilde{x})=\tilde{h}(-\tilde{x})$.

For instance,  a function with  asymptotic expansion of the form \eqref{fds} (i.e. belonging to $[\tilde{F}_n]$) that admits a dS solution $(H_0,c)=(H_0,nH_0)=\left(\displaystyle\frac{\tilde{x}_0}{\sqrt{\alpha'}},\displaystyle\frac{n\tilde{x}_0}{\sqrt{\alpha'}}\right)$ with constant dilaton for any $\tilde{x}_0$, can always be constructed by adding to $\tilde{F}_n(\tilde{x})$ a subdominant function $\tilde{h}(\tilde{x})$ such as
\be
\tilde{h}(\tilde{x})=\chi_0 \left(e^{-\frac{\chi_1}{2\tilde{x}^2}}-e^{-\frac{\chi_1}{\tilde{x}^2}}\right)-\chi_2~e^{-\frac{1}{2\tilde{x}^2}}\, ,
\ee with $\chi_0,\chi_2\in \mathbb{R}$, $\chi_1>0$. In fact, given a certain $\tilde{F}_n\in [\tilde{F}_n]$ and  $\tilde{x}_0>0$,  the $\chi_i$ can be chosen so that $\tilde{F}_n'(\tilde{x}_0)+\tilde{h}'(\tilde{x}_0)=0$ and $\tilde{F}_n(\tilde{x}_0)+\tilde{h}(\tilde{x}_0)=n^2\tilde{x}_0^2$. 
E.g. take $\chi_1$ as 
\be
 e^{-\frac{\chi_1}{2\tilde{x}_0^2}}=\frac12\iff \chi_1=\chi_1(\tilde{x}_0)=2\tilde{x}_0^2~\ln(2)>0~\, ,
\ee
$\chi_2$ as
\be
\tilde{h}'(\tilde{x}_0)=-\tilde{F}_n'(\tilde{x}_0)\iff \chi_2=\chi_2(\tilde{x}_0)=\tilde{x}_0^3~e^{\frac1{2\tilde{x}_0^2}}~\tilde{F}_n'(\tilde{x}_0) \, ,
\ee
and finally $\chi_0$ as
\be
\tilde{h}(\tilde{x}_0)=n^2\tilde{x}_0^2-\tilde{F}_n(\tilde{x}_0) \iff   \chi_0=\chi_0(\tilde{x}_0)=4~\left(n^2\tilde{x}_0^2-\tilde{F}_n(\tilde{x}_0)+\tilde{x}_0^3~\tilde{F}_n'(\tilde{x}_0)\right)\, ,
\ee
where we used the particular expressions $\chi_{1,2}(\tilde{x}_0)$ and  isolated $\chi_0$, expressing it in terms of $\tilde{x}_0$.  Therefore, the theory described non-perturbatively by $\tilde{F}_n(\tilde{x})+\tilde{h}(\tilde{x}) $  admits a dS solution $(H_0,c)=\left(\displaystyle\frac{\tilde{x}_0}{\sqrt{\alpha'}},\displaystyle\frac{n\tilde{x}_0}{\sqrt{\alpha'}}\right)$ with constant dilaton, that is stable and expanding in both frames.

Consequently,  the knowledge of the asymptotic expansion of the theory (i.e. the coefficients) is not enough to determine if it admits dS solutions of the form \eqref{eqsimpleds} or not.  Non-perturbative information is necessary, which seems to make sense since the accessible dS solutions  are non-perturbative.

In particular, if  the Lagrangian is an analytic function,  the asymptotic expansion must have a radius of convergence greater than zero. When choosing one function of the equivalence class $[\tilde{F}_h]$  equal to the convergent series in a neighborhood of zero, one is implicitly imposing non-perturbative information, since now perturbatively equivalent functions can be distinguished. Hence a subdominant function  cannot be freely added because it will break the analytic character of the Lagrangian. In principle, there seems to be no reason to assume that the Lagrangian is analytic, especially in classical string theory, which is constructed  perturbatively.

\section{Conclusions}\label{sec:conclu}

In this paper we have examined the field equations of the $\alpha'$-complete cosmology introduced in \cite{hohm}. Assuming a rather general ansatz for the fields, we determined the conditions to obtain non-perturbative dS solutions  in the string frame, and also in the Einstein frame provided the dilaton is constant. These solutions arise in the sector of vanishing Noether charge ($\mathcal{Q}=\mathbf{0}$).
We found
isotropic and anisotropic dS vacua  in $n\leq d$ spatial dimensions, with non-vanishing and vanishing $b$-field, respectively, and determined their stability. In particular, the stable and unstable dS solutions with constant dilaton are new in the context of $\alpha'$-complete cosmology, and might provide interesting implications and interpretations. Metrics with bounded scale factors can also be obtained when the $b$-field is turned on in the $\mathcal{Q}\neq\mathbf{0}$ sector.

The procedure to obtain non-perturbative dS solutions, and in particular those with constant dilaton that are proper dS geometries in the Einstein frame, was summarized in section \ref{sec:caveats}, where we further discussed their non-perturbative character. We argued that even if the complete asymptotic expansion of the theory is known, non-perturbative information  is necessary to determine if the theory admits non-perturbative dS solutions.
 Otherwise,  a subdominant function giving rise to such solutions  can always be constructed.

We conclude with some open problems and interesting  directions to continue this research.

While we have shown that the space of duality invariant cosmologies contains theories with non-perturbative dS vacua
as well as other interesting solutions, arguably an important issue is
 to determine whether  the string landscape features this type of  vacua. 
In this sense, 
the amazing achievements of  the  double-copy constructions of  all massless tree-level
amplitudes of bosonic and heterotic strings   are encouraging, as they not only seem capable of determining the full classical perturbative expansion, but also suggest a connection to  non-perturbative aspects of string theory \cite{schl} (see also \cite{mina}). Likewise, alternative constructions based on duality symmetry, such as double field theory \cite{hz} (see the reviews \cite{reviews}), have  made substantial progress in the understanding of the structure of the higher-derivative terms \cite{diego}. Establishing the precise connection between the string $\alpha'$-expansion
and the functions $F_n(H)$ in \eqref{fds} or $F_h(h_1,\cdots, h_d)$ in \eqref{efeh}  is a relevant problem to address in order to fill this gap.  

Another important question  in this direction is to establish if the no-go theorem of \cite{kutasov} applies correctly in the $\alpha'$-complete cosmology context. Under certain assumptions, the theorem rules out worldsheet constructions of dS$_n$ space-times with $n\ge 4$ in  heterotic and type II strings (without RR fluxes), and it captures all perturbative and non-perturbative $\alpha'$-corrections.
If it applies,  it would then follow that classical string theory is not one of the points in the theory space of duality covariant theories that admit non-perturbative dS solutions\footnote{We thank S. Sethi and O. Hohm for a discussion on this point.}, i.e. the function $F_n(H)$ (or $F_h(h_1,\cdots, h_d)$) that describes the string low-energy effective Lagrangian would not admit a solution of the form \eqref{eqsimpleds} (or \eqref{eqgenansatzds}). It would be interesting to understand if the subtle continuation from  Euclidean to Lorentzian signature provides a way to evade the no-go theorem.

The construction of explicit phenomenological models is another subject that deserves further examination. The higher-derivative corrections have been identified as important elements in the generation of accelerated expansion. Merged with additional effects, such as a scalar field in the Geometric Inflation scenario  \cite{edelstein} or spacetime filling KK monopoles  \cite{giuseppe}, the higher-curvature terms play a central role. From this perspective, the possible consequences  
that may result from  the solutions of the $\alpha'$-complete cosmology for model building, are worth exploring. 

For instance, it would be interesting to find bouncing cosmologies \cite{bouncing}, or new anisotropic cosmologies that resolve the Big-Bang singularity, which may include the $b$-field, thus extending \cite{non-singular} to more realistic scenarios. Another natural follow-up to our work would be to work out  the generalized ansatz  including matter, along the steps proposed in \cite{branden}. This would allow to examine interactions between matter and the $b$-field, also including more general diagonal metrics.

A   possible mechanism  for decompactification of $n=3$ spatial dimensions   was considered  in \cite{branden2},  in the spirit of the String Gas Cosmology \cite{branvafa}. This was realized assuming one dynamical and one static scale factor together with  the annihilation of winding modes  in $n$ spatial dimensions (represented by matter that verifies a certain equation of state) and their presence in the remaining $d-n$ spatial dimensions. It was shown that this model solves the size and horizon problems of Standard Big Bang cosmology if the initial value of the dilaton is sufficiently small, and also that it is compatible with the Transplanckian Censorship Conjecture \cite{branden3}, which exhibits its phenomenological relevance. A more detailed understanding of
the transitions among the different stages of
the universe modelled in \cite{branden2} (in particular of the decompactification process itself) could be gained employing the
geometries with two dynamical scale factors obtained in the previous sections.   The interaction with the $b$-field might also play an interesting role. For example, it  might supply a tool to  confine the expansion of the internal dimensions,  since the scale factor could be bounded when the $b$-field is turned on.

Proposing more general ansatze is another line of future research that might give rise to qualitatively new phenomena with potential  cosmological applications. The addition of gauge fields could also be a source of further surprises.

Finally,  the analysis of section \ref{sec:feq} can be easily extended to the isotropic Anti-dS solutions obtained in \cite{ads}, in which the fields only depend on one spatial coordinate $x$ instead of the time coordinate. More precisely, new stable and unstable non-perturbative Anti-dS solutions can be obtained with $\partial_x\Phi=-\bar{c}\neq 0,~\bar{F}(\bar{H}_0)=\bar{c}^2,~\bar{F}'(\bar{H}_0)=0$ (see \cite{ads} for definitions), and those that verify $\bar{c}=d\bar{H}_0$ have constant dilaton, thus being AdS in both the string and Einstein frames. These solutions might also provide useful applications. Moreover, the ansatz of section \ref{sec:feq} with static directions, or the general ansatz of section \ref{sec:matrixds2},  could be  worked out  in this case, including anisotropic metrics or non-vanishing $b$-field, and further lead to new non-perturbative AdS solutions.

\subsection*{Acknowledgements} We would like to thank Robert Brandenberger, Guilherme Franzmann, Olaf Hohm, Diego Marqu\'es  and Savdeep Sethi for useful comments. This work was partially supported
by  PIP-CONICET- 11220150100559CO, UBACyT 2018-2021 and  ANPCyT- PICT-2016-1358 (2017-2020).

\appendix

\section{Equations of motion in the generalized ansatz}\label{app:geneom}

In this appendix we work out the details of the procedure to obtain the equations of motion in the generalized ansatz of matrices $g,\mathcal{D}g,\mathcal{D}b$ that commute among each other. In this case, $g^{-1},\mathcal{D}(g^{-1})$ also commute with them,  and 
the matrix $(\mathcal{DS})^2$ takes the form 
\begin{equation}
\left(\mathcal{DS}\right)^2=\begin{pmatrix}A&[b,A]\\ 0&A\end{pmatrix}\, ,
\end{equation}
with $A\equiv -g^{-2}(\mathcal{D}g)^2+g^{-2}(\mathcal{D}b)^2$.
It is not hard to show by induction on $m\in\mathbb{N}$ that:
\begin{equation}
\left(\mathcal{DS}\right)^{2m}=\begin{pmatrix}A^m&[b,A^m]\\ 0&A^m\end{pmatrix}\, .
\end{equation}
In particular, $\tr[(\mathcal{DS})^{2m}]=2~\tr(A^m)$, and hence
\begin{equation}
\mathcal{F(DS)}=-2c_1\text{tr}(A)-\displaystyle\sum_{k=2}^\infty~\alpha'\,^{k-1}~\displaystyle\sum_{P\in~\text{Part}(k,2)}2^{|P|}c_{k,P}\prod_{m\in P}\text{tr}(A^{m})\equiv \mathcal{F}_a(A) \, .
\end{equation}
In addition, from \eqref{eqfprimeds} we can write
\begin{equation}\label{eqfds-fa}
\mathcal{F'(DS)}=\mathcal{DS}\cdot~\begin{pmatrix}\mathcal{F}_a'(A)&[b,\mathcal{F}_a'(A)]\\ 0&\mathcal{F}_a'(A)\end{pmatrix}\, ,
\end{equation}
and then, it is easy to see that $\tr(\mathcal{DS~F}'(\mathcal{DS}))=2\tr(A\mathcal{F}_a'(A))$. Thus, the simplest equations of motion \eqref{eomfn} and \eqref{eomfphi} turn out to be:
\begin{subequations}
\begin{align}
E_n&=(\mathcal{D}\Phi)^2-\mathcal{F}_a(A)+2\tr(A\mathcal{F}_a'(A))=0\, ,\\ 
E_\Phi+E_n&=2\mathcal{D}^2\Phi+2\tr(A\mathcal{F}_b'(A))=0\, .
\end{align}
\end{subequations}

In order to compute the equation of motion for $\mathcal{S}$ variations, or equivalently the conservation of $\mathcal{Q}$, we take the product
\be
\mathcal{S~DS}=\begin{pmatrix} g\mathcal{D}(g^{-1})+bg^{-1}(\mathcal{D}b)g^{-1}&-\mathcal{D}b-b\mathcal{D}(g^{-1})g-g\mathcal{D}(g^{-1})b-bg^{-1}(\mathcal{D}b)g^{-1}b\\
g^{-1}(\mathcal{D}b)g^{-1}&g^{-1}\mathcal{D}(g)-g^{-1}(\mathcal{D}b)g^{-1}b
\end{pmatrix}\\
\ee
This expression is absolutely general. Now, imposing the generalized ansatz and taking $g$ to be diagonal and $b(t_0)=\mathbf{0}$ without loss of generality, so that $b$ commutes with $g,g^{-1},\mathcal{D}(g^{-1}),\mathcal{D}g$ (it may not commute with $\mathcal{D}b$), this expression reduces to
\be
\begin{split}\label{eqsds}
&\mathcal{S~DS}=\begin{pmatrix} -g^{-1}\mathcal{D}g+g^{-2}b(\mathcal{D}b)&-\mathcal{D}b+2(g^{-1}\mathcal{D}g)b-g^{-2}b(\mathcal{D}b)b\\
g^{-2}(\mathcal{D}b)&g^{-1}\mathcal{D}(g)-g^{-2}(\mathcal{D}b)b
\end{pmatrix}\\
\end{split}
\ee

Therefore, in this generalized ansatz the conservation of the Noether charge takes the form:
\be\label{eqqgeneralizedansatz}
-\displaystyle\frac{\mathcal{Q}}{2}=e^{-\Phi}\mathcal{S}\mathcal{F}'(\mathcal{DS})\equiv \begin{pmatrix}{\mathcal{Q}}_{3}-{\mathcal{Q}}_2&{\mathcal{Q}}_{4}\\{\mathcal{Q}}_1&{\mathcal{Q}}_{3}+{\mathcal{Q}}_2\end{pmatrix}\, ,
\ee
or equivalently, computing each block of $\mathcal{S}\mathcal{F}'(\mathcal{DS})$ using \eqref{eqfds-fa} and \eqref{eqsds}:
\begin{subequations}
\begin{align}
&{\mathcal{Q}}_{3}-{\mathcal{Q}}_2=e^{-\Phi}~(-g^{-1}\mathcal{D}g+b~g^{-2}(\mathcal{D}b))\mathcal{F}_a'(A)\, ,\\
&{\mathcal{Q}}_{4}=e^{-\Phi}~\left((-\mathcal{D}b+bg^{-1}\mathcal{D}g)\mathcal{F}_a'(A)-(-g^{-1}\mathcal{D}g+bg^{-2}(\mathcal{D}b))\mathcal{F}_a'(A)b\right)\, ,\\
&{\mathcal{Q}}_1=e^{-\Phi}~g^{-2}(\mathcal{D}b)\mathcal{F}_a'(A)\, ,\\
&{\mathcal{Q}}_{3}+{\mathcal{Q}}_2=e^{-\Phi}~\left(g^{-1}\mathcal{D}(g)\mathcal{F}_a'(A)-g^{-2}(\mathcal{D}b)\mathcal{F}_a'(A)b\right)\, .
\end{align}
\end{subequations}
The condition  $\mathcal{Q}\in \mathfrak{so}(d,d)$ means that   ${\mathcal{Q}}_1,{\mathcal{Q}}_{3},{\mathcal{Q}}_{4}$ must be antisymmetric  and ${\mathcal{Q}}_2$ symmetric.

The first and second equations can be used to rewrite the previous system of equations as
\begin{subequations}
\begin{align}
{\mathcal{Q}}_1&=e^{-\Phi}~g^{-2}(\mathcal{D}b)\mathcal{F}_a'(A)\, , \qquad
{\mathcal{Q}}_2=e^{-\Phi}~g^{-1}\mathcal{D}(g)\mathcal{F}_a'(A)-\frac12\{b,{\mathcal{Q}}_1\}\, ,\\
2{\mathcal{Q}}_{3}&=[b,{\mathcal{Q}}_{1}]\, , \qquad \qquad
{\mathcal{Q}}_{4}=-g^2~{\mathcal{Q}}_1+b\left({\mathcal{Q}}_2+\frac12\{b,{\mathcal{Q}}_1\}\right)-({\mathcal{Q}}_{3}-{\mathcal{Q}}_2)b
\end{align}
\end{subequations}
The first equation implies that ${\mathcal{Q}}_1$ commutes with $g,g^{-1},\mathcal{D}g,\mathcal{D}b$ for all times. Then, the third equation  is
 trivially  verified for ${\mathcal{Q}}_{3}=\mathbf{0}$, since  $b(t_0)=\mathbf{0}$, and thus
\be\nonumber
\mathcal{D}\left([b(t),{\mathcal{Q}}_1]\right)=[\mathcal{D}b(t),{\mathcal{Q}}_1]=0 \implies 2{\mathcal{Q}}_3=[b(t),{\mathcal{Q}}_1]=[b(t_0),{\mathcal{Q}}_1]=\mathbf{0} \, .\ee

Hence, the equations are also equivalent to:
\be
{\mathcal{Q}}_1=e^{-\Phi}~g^{-2}(\mathcal{D}b)\mathcal{F}_a'(A)\, ,\ \
{\mathcal{Q}}_2=e^{-\Phi}~g^{-1}\mathcal{D}(g)\mathcal{F}_a'(A)-b{\mathcal{Q}}_1\, , \ \
{\mathcal{Q}}_{4}=-g^2~{\mathcal{Q}}_1+\{b,{\mathcal{Q}}_2\}+b^2{\mathcal{Q}}_1\nonumber
\ee
In particular, we may evaluate the third equation in $t_0$ considering $b(t_0)=\mathbf{0}$ and obtain
\be
{\mathcal{Q}}_{4}=-g^2(t_0)~{\mathcal{Q}}_1 \, .
\ee
Notice that the third equation is equivalent to
\be\nonumber
\mathbf{0}=\mathcal{D}{\mathcal{Q}}_{4}=-2g\mathcal{D}g~{\mathcal{Q}}_1+\{\mathcal{D}b,{\mathcal{Q}}_2\}+\{\mathcal{D}b,b{\mathcal{Q}}_1\}=-2g\mathcal{D}g~{\mathcal{Q}}_1+2g\mathcal{D}g~{\mathcal{Q}}_1
\ee
where we used   the fact that $\mathcal{D}b$ commutes with $g^{-1},\mathcal{D}g,A$.
Therefore,  it  is automatically verified from  the first two, with an integration constant ${\mathcal{Q}}_{4}$.
Hence, the equation of motion for $\mathcal{S}$ variations reduces to 
\begin{subequations}
\begin{align}
{\mathcal{Q}}_1&=e^{-\Phi}~g^{-2}(\mathcal{D}b)\mathcal{F}_a'(A)\label{eqa3second}\\
{\mathcal{Q}}_2&=e^{-\Phi}~g^{-1}\mathcal{D}(g)\mathcal{F}_a'(A)-b{\mathcal{Q}}_1=e^{-\Phi}~(g^{-1}\mathcal{D}(g)-g^{-2}b\mathcal{D}b)\mathcal{F}_a'(A)\label{eqc1second}
\end{align}
\end{subequations}
${\mathcal{Q}}_1$ and ${\mathcal{Q}}_2$ are block diagonal with the same blocks as $\mathcal{D}b$, since $A,\mathcal{D}b,b,g^{-2},\mathcal{D}g$ are block diagonal matrices, and hence they commute with $g,g^{-1},\mathcal{D}g$.
They can also be expressed without the integration constants ${\mathcal{Q}}_1,{\mathcal{Q}}_2$, as:
\begin{subequations}
\begin{align}
\mathbf{0}&=\mathcal{D}\left(e^{-\Phi}~g^{-2}(\mathcal{D}b)\mathcal{F}_a'(A)\right)\label{eqa3der}\\
\mathbf{0}&=\mathcal{D}\left(e^{-\Phi}~g^{-1}\mathcal{D}(g)\mathcal{F}_a'(A)\right)-\underbrace{e^{-\Phi}~g^{-2}(\mathcal{D}b)^2\mathcal{F}_a'(A)}_{\displaystyle=\mathcal{D}b~{\mathcal{Q}}_1}\label{eqc1der}
\end{align}
\end{subequations}
In summary, the equations of motion are:
\begin{subequations}
\begin{align}
&{\mathcal{Q}}_1=e^{-\Phi}~g^{-2}(\mathcal{D}b)\mathcal{F}_a'(A)=\text{constant} \iff \mathbf{0}=\mathcal{D}\left(e^{-\Phi}~g^{-2}(\mathcal{D}b)\mathcal{F}_a'(A)\right)\label{eqa3-4}\\
&{\mathcal{Q}}_2=e^{-\Phi}~g^{-1}\mathcal{D}(g)\mathcal{F}_a'(A)-b{\mathcal{Q}}_1=\text{constant}\iff \mathbf{0}=\mathcal{D}\left(e^{-\Phi}~g^{-1}\mathcal{D}(g)\mathcal{F}_a'(A)\right)-\mathcal{D}b~{\mathcal{Q}}_1\label{eqc1-4}\\
&E_n=(\mathcal{D}\Phi)^2-\mathcal{F}_a(A)+2\tr(A\mathcal{F}_a'(A))=0\\ 
&E_\Phi+E_n=2\mathcal{D}^2\Phi+2\tr(A\mathcal{F}_a'(A))=0
\end{align}
\end{subequations}

\subsection{Diagonalized equations of motion}\label{subapp:diag}

Since $\mathcal{D}b$ and ${\mathcal{Q}}_1$ are real antisymmetric matrices (hence anti-hermitian) that commute, they can be simultaneously diagonalized with a complex unitary matrix $U(t)$:
\be\label{eqdbdiag-app}
\mathcal{D}b(t)=U(t)(iD_b(t))U^{-1}(t)\, ,\qquad
{\mathcal{Q}}_1=U(t)(iD_1)U^{-1}(t)\, ,
\ee
with $D_b$ and $D_{1}$ real diagonal matrices. Likewise, since $\mathcal{D}b$ and ${\mathcal{Q}}_1$ are block diagonal, the unitary matrices $U(t),U^{-1}(t)$ are block diagonal with blocks of equal scale factor, hence they commute with $g,g^{-1},\mathcal{D}g$. In principle, they depend on time.

Moreover, this implies that $A$ is expressed in this basis as:
\be
A=U(t)(-g^{-2}(\mathcal{D}g)^2-g^{-2}D_b^2)U^{-1}(t)\equiv U(t)(-4D^2)U^{-1}(t)
\ee
where we defined the diagonal matrix $D^2$ with non-negative real elements, recalling that $g,g^{-1},\mathcal{D}g$ are diagonal matrices. Notice that $\mathcal{F}_a'(A)=U(t)~\mathcal{F}_a'(-4D^2)~U^{-1}(t)$.

The equations \eqref{eqa3-4} and \eqref{eqc1-4} then take the form: 
\begin{subequations}
\begin{align}
-i{\mathcal{Q}}_1&=U(t)D_{1}U^{-1}(t)=U(t)~e^{-\Phi}~g^{-2}(D_b)\mathcal{F}_a'(-4D^2)~U^{-1}(t)=\text{ constant}\label{eqa3-5}\\
\mathbf{0}&=\mathcal{D}\left(U(t)~e^{-\Phi}~g^{-1}\mathcal{D}(g)\mathcal{F}_a'(-4D^2)~U^{-1}(t)\right)+U(t)~D_b~D_{1}~U^{-1}(t)\label{eqc1-5}
\end{align}
\end{subequations}

From \eqref{eqa3-5} we see that:
\be\label{eqd3-1}
D_{1}=e^{-\Phi}~g^{-2}(D_b)\mathcal{F}_a'(-4D^2)=\text{ constant}
\ee
since the elements of $D_{1}$ are the eigenvalues of $-i{\mathcal{Q}}_1$, which are constant if ${\mathcal{Q}}_1$ is constant.  We still need to impose that $-i{\mathcal{Q}}_1=U(t)D_{1}U^{-1}(t)$ is constant:
\be\label{equ1}
\mathbf{0}=-i\mathcal{D}{\mathcal{Q}}_1=U(t)(\overbrace{\mathcal{D}D_{1}}^{=\mathbf{0}}+[U^{-1}\mathcal{D}U,D_{1}])U^{-1}(t)\iff [U^{-1}\mathcal{D}U,D_{1}]=\mathbf{0}
\ee
This is a condition for  $U(t)$, and then \eqref{eqa3-5} is equivalent to both \eqref{eqd3-1} and \eqref{equ1}.

Equation \eqref{eqc1-5}  may be written equivalently as
\be
\begin{split}
\mathbf{0}&=D_b~D_{1}+U^{-1}(t)~\mathcal{D}\left(U(t)~e^{-\Phi}~g^{-1}\mathcal{D}(g)\mathcal{F}_a'(-4D^2)~U^{-1}(t)\right)~U(t)\\
&=D_b~D_{1}+\mathcal{D}\left(e^{-\Phi}~g^{-1}\mathcal{D}(g)\mathcal{F}_a'(-4D^2)\right)+[U^{-1}\mathcal{D}U,e^{-\Phi}~g^{-1}\mathcal{D}(g)\mathcal{F}_a'(-4D^2)]\\
\end{split}
\ee
Notice that the first two terms of the last line are diagonal matrices, while the last term is not. Moreover, the latter is of the form $[U^{-1}\mathcal{D}U,D^G]$ where $D^{G}$ is a generic diagonal matrix, hence it has matrix elements of the form $[U^{-1}\mathcal{D}U,D^G]_{ij}=(U^{-1}\mathcal{D}U)_{ij}(D_j^G-D_i^G)$, with null diagonal elements. Therefore, we may project the diagonal and non-diagonal elements of \eqref{eqc1-5} as
\be
\begin{split}
\mathbf{0}&=D_b~D_{1}+\mathcal{D}\left(e^{-\Phi}~g^{-1}\mathcal{D}(g)\mathcal{F}_a'(-4D^2)\right)\\
\mathbf{0}&=\mathcal{D}(g)~[U^{-1}\mathcal{D}U,\mathcal{F}_a'(-4D^2)]
\end{split}
\ee
where we used that $U(t),U^{-1}(t)$ are block diagonal matrices and $e^{-\Phi}g^{-1}$ is invertible.

Finally,  the equations of motion take the form
\begin{subequations}\label{eqsmotion-6}
\begin{align}
D_{1}&=e^{-\Phi}~g^{-2}(D_b)\mathcal{F}_a'(-4D^2)=\text{ constant}\label{eqa3-6}\\
\mathbf{0}&=D_b~D_{1}+\mathcal{D}\left(e^{-\Phi}~g^{-1}\mathcal{D}(g)\mathcal{F}_a'(-4D^2)\right)\label{eqc1-6}\\
E_n&=(\mathcal{D}\Phi)^2-\mathcal{F}_a(-4D^2)+2\tr((-4D^2)~\mathcal{F}_a'(-4D^2))=0\\ 
E_\Phi+E_n&=2\mathcal{D}^2\Phi+2\tr((-4D^2)~\mathcal{F}_a'(-4D^2))=0\\
\mathbf{0}&=[U^{-1}\mathcal{D}U,D_1]=\mathcal{D}(g)~[U^{-1}\mathcal{D}U,\mathcal{F}_a'(-4D^2)]\label{equnitaryu}
\end{align}
\end{subequations}

\section{Constant \texorpdfstring{$h_i$}{1} and \texorpdfstring{$\mathcal{Q}=\mathbf{0}$}{1}}\label{app:qne0}

In this appendix we show that $h_i=$ constant implies  $\mathcal{Q}=\mathbf{0}$. This is the reason why a vanishing Noether charge is  quite a rich sector to find non-perturbative dS solutions.

The equations that determine the dynamics of $h_i(t)=h^{(r)}_{\alpha}(t)$ are:
\be
e^{-\Phi}\frac{\partial F_h}{\partial h_i}= q_i=\text{ constant} \iff \nabla F_h=e^\Phi~\vec{q}
\ee
for each $1\leq i\leq d$. 
Defining $q\equiv \|\vec{q}\|$ and the dimensionless vector $\vec{w}\equiv \displaystyle\frac{\vec{q}}{q}$ of norm $\|\vec{w}\|=1$, only in the case 
\be
q\neq 0\iff \vec{q}\neq \vec{0}\iff \exists~ r,\alpha \text{ such that } Q^{(r)}_{\alpha}\neq 0\text{ or }Q'\,^{(r)}_{\alpha}\neq 0\iff\mathcal{Q}\neq \mathbf{0} \, , 
\ee
one can perform a change of variables of the form
\be
\vec{h}=(h_1,h_2,\cdots,h_d)\to \vec{X}=(X_1,X_2,\cdots,X_d)=W\vec{h}
\ee
where $W$ is a constant real orthogonal matrix (hence $\vec{h}=W^{-1}\vec{X}=W^{t}\vec{X}$) such that its first row is equal to the unitary and dimensionless vector $\vec{w}$ (i.e. $w_i=W_{1i}=(W^t)_{i1}$), and obviously the rest of the rows are orthogonal to $\vec{w}$. Then the partial derivatives become:
\begin{equation}\label{eqfequisi}
\begin{split}
&\displaystyle\frac{\partial F_{h}}{\partial X_1}=\displaystyle\sum_{i=1}^d \frac{\partial h_i}{\partial X_1}\frac{\partial F_{h}}{\partial h_i}=q~e^\Phi~\sum_{i=1}^d (W^t)_{i 1}w_i=q~e^\Phi~\sum_{i=1}^d w_{i}w_i=q~e^\Phi~\|\vec{w}\|^2= q~e^\Phi\neq 0 \\
&\displaystyle\frac{\partial F_{h}}{\partial X_\chi}=\displaystyle\sum_{i=1}^d \frac{\partial h_i}{\partial X_\chi}\frac{\partial F_{h}}{\partial h_i}=q~e^\Phi~\sum_{i=1}^d (W^t)_{i \chi}w_i=q~e^\Phi~\sum_{i=1}^d (W)_{\chi i}w_i=0~~\text{ for  }~ 2\leq \chi\leq d\, .\\
\end{split}
\end{equation}

Since  some of the $h_i$ may have equal modulus (e.g. if $\beta_\alpha^{(r)}\neq 0$, there is another $\beta_{\alpha'}^{(r)}=-\beta_\alpha^{(r)}$, hence $|h_{\alpha'}^{(r)}|=|h_{\alpha}^{(r)}|$; or if $\beta_\alpha^{(r)}= 0$ for various values of $\alpha$ and the same $r$), and some $h_i$ may be equal to $0$, various linear combinations of the $h_i$ may be trivial (e.g. $h_{\alpha'}^{(r)}\pm h_{\alpha}^{(r)}=0$ and certain $X_\chi=0$, then trivially $\displaystyle\frac{\partial F_h}{\partial X_\chi}=0$, or $\displaystyle\frac{\partial F_h}{\partial X_\chi}=0$ for $X_\chi=h_i=0$). However, if there are at most $m$ values of $h_i$ with distinct non-zero modulus, there will be in principle $m$ non-trivial linear combinations $X_i$ and $d-m$ trivial linear combinations $X_\chi=0$, for which $\displaystyle\frac{\partial F_h}{\partial X_\chi}=0$.

If  every $h_i=h^{(r')}_{\alpha'}=$  constant, or equivalently every $X_i=$  constant, then  from \eqref{eqEnFh},  $\mathcal{D}\Phi=-c=$constant $\in \mathbb{R}$  follows, which in turn implies from \eqref{eqEphiFh} that 
\be
\begin{split}
0&=\displaystyle\sum_{i=1}^d h_i\displaystyle\frac{\partial F_h}{\partial h_i}=\displaystyle\sum_{i=1}^d \displaystyle\sum_{j=1}^d\displaystyle\sum_{k=1}^d (W^t)_{ij}X_j\displaystyle\frac{\partial X_k}{\partial h_i}\frac{\partial F_h}{\partial X_k}=\displaystyle\sum_{i=1}^d \displaystyle\sum_{j=1}^d\displaystyle\sum_{k=1}^d (W^t)_{ij}X_jW_{ki}\frac{\partial F_h}{\partial X_k}\\
&=\displaystyle\sum_{j=1}^d \displaystyle\sum_{k=1}^d X_j \frac{\partial F_h}{\partial X_k}(WW^t)_{kj}=\displaystyle\sum_{j=1}^dX_j \frac{\partial F_h}{\partial X_j} =X_1\frac{\partial F_h}{\partial X_1}\, ,
\end{split}
\ee
where we used $(WW^t)_{kj}=\delta_{kj}$ because $W$ is orthogonal, and $\displaystyle\frac{\partial F_h}{\partial X_\chi}=0$ for $\chi\geq 2$.

Since the solution $X_1=0$ is contained in $\displaystyle\frac{\partial F_h}{\partial X_1}=0$, this equation is equivalent to:
\be\label{eqqequal0hconst}
\frac{\partial F_h}{\partial X_1}=0 \iff q=0 \iff \mathcal{Q}=\mathbf{0}\, .
\ee
Hence, we see that imposing $h_i=$constant for all $1\leq i\leq d$, necessarily implies that $\mathcal{Q}=\mathbf{0}$.

Conversely, the simplest way of constructing $\mathcal{Q}=\mathbf{0}$ solutions, which verify $\nabla F_h=\vec{0}$, is to impose that every $h_i$ is constant. Otherwise, if there were some time-dependent $h_i(t)$, the equation $\nabla F_h=\vec{0}$ would hold in a certain open neighborhood of $h_i(t_0)$, and then the function $F_h$ would not depend on $h_i$ (which may be a non-trivial condition for the asymptotic expansion \eqref{efeh} of $F_h$).

\section{Caveats on generalized non-perturbative dS solutions}\label{app:caveats}

In this appendix we extend the discussion on non-perturbative dS solutions presented in section \ref{sec:caveats} to the case of the generalized ansatz of commuting matrices.

The procedure to find non-perturbative dS solutions immediately extends to  the generalized ansatz,  considering that in this case one has to calculate the roots $\vec{h}_0=(h_{0,1},\cdots,h_{0,d})\neq \vec{0}$ of $\nabla F_h$ in the step 1. Given a root $\vec{h}_0$, one can always construct $2^d-1$ different new roots $\vec{h}_0'=(\pm h_{0,1},\cdots, \pm h_{0,d})$ since they also verify $\displaystyle\frac{\partial F_h}{\partial h_i}=0$ for all $i$. Hence, to have an expanding cosmology (in the string frame) in certain spatial directions,  one should choose only the roots with positive Hubble parameter in such spatial directions. Moreover,  an expanding or contracting cosmology in the remaining spatial directions can be chosen if they are not static. The subsequent steps are trivially generalized.

The construction of a subdominant function can also be done in the generalized ansatz. We first define the dimensionless function $\tilde{F}_h(\vec{\tilde{x}})\equiv \alpha'F_h(\vec{h})$ of the dimensionless variables $\tilde{x}_i\equiv \sqrt{\alpha'}~h_i$. In order to ensure that the multi-variable subdominant function $\tilde{h}(\vec{\tilde{x}})$ is time-reversal and duality invariant (considering that the symmetries hold non-perturbatively), we take it to depend only on the combinations $S_k\equiv\displaystyle\sum_{i=1}^d \tilde{x}_i^{2k}\propto \alpha'\,^{k}\tr(\mathcal{DS}^{2k})$. More precisely:
\be
\tilde{h}(\vec{\tilde{x}})= \tilde{h}_S(S_1,S_2,\cdots)=\chi_0\left(e^{-\frac{\chi_1}{2S_1}}-e^{-\frac{\chi_1}{S_1}}\right)+\tilde{h}_{S,1}(S_1,S_2,\cdots)\, .
\ee
Consider $\chi_1=2~\ln(2)~ S_1>0$, hence the first derivatives of the term proportional to $\chi_0$ are zero. Thus we can always choose $\chi_0\in \mathbb{R}$ such that $\tilde{h}(\vec{\tilde{x}}_0)=\tilde{c}^2-\tilde{F}_h(\vec{\tilde{x}}_0)$ for some $\tilde{c}$, which we may take such that the dilaton is constant. This $\chi_0$-term does not influence the condition of the first derivatives, consequently we only need to work it out with $\tilde{h}_{S,1}(S_1,S_2,\cdots)$.

Consider there are at most $1\leq m\leq d$ elements $\tilde{x}_{0,I}$ in the chosen vector $\vec{\tilde{x}}_0$ with distinct non-zero modulus (i.e. $|\tilde{x}_{0,I}|\neq0$ and $I\neq I'\implies |\tilde{x}_{0,I}|\neq |\tilde{x}_{0,I'}|$ for $I,I'$ taking $m$ possible values). In addition, the remaining $d-m$ elements $\tilde{x}_{0,i}$ must either verify $|\tilde{x}_{0,i}|=|\tilde{x}_{0,I}|$ for some $I$, or $\tilde{x}_{0,i}=0$. Then, for any duality and time-reversal invariant function $\tilde{F}_h+\tilde{h}$, we only need the $m$ partial derivatives $\displaystyle\frac{\partial (\tilde{F}_h+\tilde{h})}{\partial \tilde{x}_I}(\vec{\tilde{x}}_0)$ to compute its gradient, because the remaining $d-m$ partial derivatives with respect to $\tilde{x}_i$ either verify $\displaystyle\frac{\partial (\tilde{F}_h+\tilde{h})}{\partial \tilde{x}_i}(\vec{\tilde{x}}_0)=\pm\displaystyle\frac{\partial (\tilde{F}_h+\tilde{h})}{\partial \tilde{x}_I}(\vec{\tilde{x}}_0)$ for an index $I$ (with $\pm=\sign(\tilde{x}_{0,i})\sign(\tilde{x}_{0,I})$), or trivially verify $\displaystyle\frac{\partial (\tilde{F}_h+\tilde{h})}{\partial \tilde{x}_i}(\vec{\tilde{x}}_0)=0$ if $\tilde{x}_{0,i}=0$. In particular, to impose $\displaystyle\frac{\partial (\tilde{F}_h+\tilde{h})}{\partial \tilde{x}_i}(\vec{\tilde{x}}_0)=0$ for every index $1\leq i\leq d$ (i.e. $\nabla (\tilde{F}_h+\tilde{h})=\vec{0}$) is equivalent to impose $\displaystyle\frac{\partial (\tilde{F}_h+\tilde{h})}{\partial \tilde{x}_I}(\vec{\tilde{x}}_0)=0$ only for the $m$ indexes $I$. For simplicity, we order the indexes in such a way that $1\leq I\leq m$ and the remaining $m+1\leq i\leq d$.

We may choose $\tilde{h}_{S,1}(S_1,S_2,\cdots)=\displaystyle\sum_{J=1}^m \chi_J ~e^{-1/S_J}$. Therefore we would like to choose $\chi_J\in \mathbb{R}$ such that $\displaystyle\frac{\partial \tilde{h}}{\partial \tilde{x}_I}(\vec{\tilde{x}}_0)=\left.\displaystyle\sum_{J=1}^m \chi_J~\displaystyle\frac{2J~\tilde{x}_I^{2J-1}}{S_J^2} ~e^{-1/S_J}\right|_{\vec{\tilde{x}}_0}=-\displaystyle\frac{\partial \tilde{F}_h}{\partial \tilde{x}_I}(\vec{\tilde{x}}_0)$ for each $1\leq I\leq m$. We may write this as a matrix equation of the form $\bar{\bar{M}}\cdot\vec{\chi}=\nabla_m \tilde{F}_h(\vec{\tilde{x}}_0)$, where $\nabla_m$ only includes the first $m$ partial derivatives, and $\bar{\bar{M}}$ is a $m\times m$ matrix with elements $M_{IJ}=\left.-\displaystyle\frac{2J~\tilde{x}_{0,I}^{2J-1}}{S_J^2} ~e^{-1/S_J}\right|_{\vec{\tilde{x}}_0}$. Using that $\tilde{x}_{0,I}\neq 0$ and hence $S_J|_{\vec{\tilde{x}}_0}\neq 0$, it is easy to check that the determinant of $\bar{\bar{M}}$ is non-zero if and only if the determinant of the Vandermonde matrix $V(\tilde{x}_{0,1}^{2},\tilde{x}_{0,2}^{2},\cdots,\tilde{x}_{0,m}^{2})$ (with elements $V_{IJ}=\tilde{x}_{0,I}^{2(J-1)}$) is non-zero. Since the latter is equal to $\displaystyle\prod_{1\leq J<K\leq m}(\tilde{x}_{0,K}^2-\tilde{x}_{0,J}^2)$, and $J\neq K\implies \tilde{x}_{0,K}^2\neq\tilde{x}_{0,J}^2$,  the determinant of $V$ is non-zero, and hence the determinant of $\bar{\bar{M}}$ is also non-zero (i.e. it is invertible). This implies that  the linear system  can always be solved and the coefficients $\vec{\chi}=\bar{\bar{M}}^{-1}\cdot\nabla_m \tilde{F}_h(\vec{\tilde{x}}_0)$ be obtained.

Therefore,  a duality and time-reversal invariant subdominant function $\tilde{h}(\vec{\tilde{x}})$ can always be constructed, such that the first $m$ partial derivatives of $\tilde{F}_h+\tilde{h}$ are zero, and hence all the $d$ partial derivatives of $\tilde{F}_h+\tilde{h}$ are also zero, always evaluated in $\vec{\tilde{x}}_0$. In addition, as we previously explained, one can also impose $\tilde{F}_h(\vec{\tilde{x}}_0)+\tilde{h}(\vec{\tilde{x}}_0)=\tilde{c}^2$ for some $\tilde{c}$ (which may be taken such that the dilaton is constant), by choosing $\chi_0\in\mathbb{R}$ accordingly, extending the results of section \ref{sec:caveats}.

\end{document}